\documentclass[acmsmall,nonacm,preprint]{acmart}

\usepackage[title]{appendix}
\usepackage{array}
\usepackage{microtype}
\usepackage{rotating}
\usepackage{pdflscape}
\usepackage{geometry}
\usepackage{adjustbox}
\usepackage{lscape}
\usepackage{longtable}
\usepackage{import}
\usepackage{graphicx}
\usepackage{amsmath}
\usepackage{tabularx}
\usepackage{colortbl}
\usepackage{multirow}
\usepackage{color}
\usepackage{multicol}
\usepackage{booktabs}
\usepackage{makecell}
\usepackage{caption}
\usepackage{subcaption}
\usepackage{url}
\usepackage{algpseudocode}
\usepackage{algorithm}
\usepackage{textcomp}
\usepackage{xcolor}
\usepackage{xspace}
\usepackage{comment}
\usepackage{enumitem}
\usepackage{xfrac}
\usepackage{adjustbox}
\usepackage[most]{tcolorbox}
\usepackage{ragged2e}
\usepackage[normalem]{ulem}
\usepackage{enumitem}
\usepackage{xcolor,colortbl}
\usepackage{siunitx}
\usepackage{ifthen}
\usepackage{parcolumns}
\usepackage{ltablex}
\usepackage{pdfpages}
\newcommand*{\priority}[1]{\begin{tikzpicture}[scale=0.15]%
    \draw (0,0) circle (1);
    \ifthenelse{#1>0}{\fill[fill opacity=0.5,fill=black] (0,0) -- (90:1) arc (90:90-#1*3.6:1) -- cycle;}{}
    \end{tikzpicture}}

\usepackage{pifont}
\usepackage{colortbl}
\usepackage{ragged2e}

\definecolor{highrisk}{RGB}{250,236,231}
\definecolor{hightext}{RGB}{153,60,29}
\definecolor{midrisk}{RGB}{230,241,251}
\definecolor{midtext}{RGB}{24,95,165}
\definecolor{modrisk}{RGB}{250,238,218}
\definecolor{modtext}{RGB}{133,79,11}
\definecolor{fairrisk}{RGB}{234,243,222}
\definecolor{fairtext}{RGB}{59,109,17}
\definecolor{headerbg}{RGB}{245,245,243}
\definecolor{rowalt}{RGB}{250,250,249}

\newcommand{\highbadge}{\colorbox{highrisk}{\textcolor{hightext}{\footnotesize\textbf{High}}}}
\newcommand{\midbadge}{\colorbox{midrisk}{\textcolor{midtext}{\footnotesize\textbf{Mid}}}}
\newcommand{\modbadge}{\colorbox{modrisk}{\textcolor{modtext}{\footnotesize\textbf{Moderate}}}}
\newcommand{\fairbadge}{\colorbox{fairrisk}{\textcolor{fairtext}{\footnotesize\textbf{Fair}}}}

\usepackage{tikz}
\definecolor{darkblue}{rgb}{0.0, 0.0, 0.5}
\definecolor{lightgray}{gray}{0.9}
\definecolor{lightblue}{rgb}{0.68, 0.85, 0.90}

\newcolumntype{N}{>{\centering\arraybackslash}m{.02in}}
\newcolumntype{G}{>{\centering\arraybackslash}m{0.5in}}

\newcommand{\new}[1]{{\color{black}#1}}

\usepackage{titlesec}
\usepackage{hyperref}
\usepackage{tabulary}

\titleclass{\subsubsubsection}{straight}[\subsection]

\newcounter{subsubsubsection}[subsubsection]
\renewcommand\thesubsubsubsection{\thesubsubsection.\arabic{subsubsubsection}}

\titleformat{\subsubsubsection}
  {\normalfont\normalsize\bfseries}{\thesubsubsubsection}{1em}{}
\titlespacing*{\subsubsubsection}
{0pt}{3.25ex plus 1ex minus .2ex}{1.5ex plus .2ex}

\makeatletter
\renewcommand\paragraph{\@startsection{paragraph}{5}{\z@}%
  {3.25ex \@plus1ex \@minus.2ex}
  {-1em}
  {\normalfont\normalsize\bfseries}}
\renewcommand\subparagraph{\@startsection{subparagraph}{6}{\parindent}%
  {3.25ex \@plus1ex \@minus .2ex}%
  {-1em}%
  {\normalfont\normalsize\bfseries}}
\def\toclevel@subsubsubsection{4}
\def\toclevel@paragraph{5}
\def\toclevel@paragraph{6}
\def\l@subsubsubsection{\@dottedtocline{4}{7em}{4em}}
\def\l@paragraph{\@dottedtocline{5}{10em}{5em}}
\def\l@subparagraph{\@dottedtocline{6}{14em}{6em}}
\makeatother

\AtBeginDocument{%
  \providecommand\BibTeX{{%
    \normalfont B\kern-0.5em{\scshape i\kern-0.25em b}\kern-0.8em\TeX}}}

\setcopyright{acmcopyright}
\copyrightyear{2018}
\acmYear{2018}
\acmDOI{XXXXXXX.XXXXXXX}

\acmJournal{JACM}
\acmVolume{37}
\acmNumber{4}
\acmArticle{111}
\acmMonth{8}

\begin{document}

\title{Security Below the OS – A Security Analysis of UEFI}

\author{Priyanka Prakash Surve}
\email{surve@post.bgu.ac.il}
\orcid{0009-0001-5673-2687}
\affiliation{%
  \institution{Ben-Gurion University of the Negev}
  \city{Be'er Sheva}
  \country{Israel}
}

\author{Oleg Brodt}
\email{boleg@bgu.ac.il}

\affiliation{%
  \institution{Ben-Gurion University of the Negev}
  \city{Be'er Sheva}
  \country{Israel}
}

\author{Mark Yampolskiy}
\email{mark.yampolskiy@auburn.edu}

\affiliation{%
  \institution{Auburn University}
  \country{United States}
}

\author{Yuval Elovici}
\email{elovici@bgu.ac.il}
\orcid{0000-0002-9641-128X}
\affiliation{%
  \institution{Ben-Gurion University of the Negev}
  \city{Be'er Sheva}
  \country{Israel}
}

\author{Asaf Shabtai}
\email{shabtaia@bgu.ac.il}
\orcid{0000-0003-0630-4059}
\affiliation{%
  \institution{Ben-Gurion University of the Negev}
  \city{Be'er Sheva}
  \country{Israel}
}

\renewcommand{\shortauthors}{Surve, et al.}

\begin{abstract}
The Unified Extensible Firmware Interface (UEFI) plays a crucial role in modern computing systems, governing secure system initialization and booting.
A sharp spike in UEFI-related attacks and vulnerabilities has been observed in recent years~\new{\cite{eclypsium2024kev}}.
To respond to this alarming trend, we believe that the cybersecurity community should be equipped with the knowledge to understand the UEFI landscape, the related attacks, and potential defenses. 
Over the years UEFI has been a niche topic in cybersecurity, attracting little research and lacking a comprehensive overview. 
The knowledge that exists on the topic is scattered across official documentation, blog posts, papers, and books which are not necessarily security focused, creating a barrier to entry for cybersecurity professionals who want to learn about the topic. This paper aims to correct that problem by exploring the UEFI from a security point of view, including the UEFI landscape, the UEFI development lifecycle, distribution, supply chain, and booting process. 
We investigate a wide range of real-world UEFI attacks and potential attack vectors that could exploit vulnerabilities at the various stages of the UEFI lifecycle that we examine in this paper. 
Inspired by the MITRE ATT\&CK framework, we present a taxonomy delineating the tactics, techniques, and sub-techniques of UEFI attacks based on the commonalities and patterns observed in our examination of the attacks.
Further to outlining attacks and defenses, we perform a risk analysis to identify the techniques that pose the greatest threat to UEFI security. 
Prioritizing the most critical risks will enable future research aimed at addressing the most vulnerable aspects of the UEFI and contributing to its security.

\end{abstract}



\maketitle

\section{Introduction}
\label{sec:intro}
UEFI is located on the SPI chip of the motherboard and it is the first code that runs after the computer is switched on.
UEFI specifications define the interface between the computer hardware and the operating system (OS) on modern computer systems and provide a standardized way for initializing the computer hardware, booting the operating system, and providing runtime services~\cite {uefi_organization}.
\new{Throughout this paper, the term \textit{UEFI} refers specifically to the Unified Extensible Firmware Interface and its associated firmware. The term \textit{BIOS} is used only when referring to legacy Basic Input/Output System firmware predating UEFI, or when citing vendor or tool terminology that uses BIOS to refer to the firmware flash chip generically, as is common in industry practice.}

UEFI offers enhanced security features as compared to traditional Basic Input/Output Systems (BIOS), but despite continuous enhancements and advancements in UEFI security it is increasingly becoming a lucrative target for cyber threat actors~\cite{cisa2023call}.
This trend is largely due to the complexity of (i) the UEFI supply chain, where multiple actors, including hardware manufacturers, software developers, and end-users play an important role at different stages of the UEFI development and maintenance life cycle (yet their responsibilities often overlap and extend beyond a single phase)~\cite{blackhat_uefi_safeguarding}; and (ii) the architecture of UEFI itself, which is designed to provide a more flexible and feature-rich environment for booting systems than the legacy BIOS, through a phased booting process where control is handed off from one phase to another, and through the support of multiple drivers and functionalities.
The complexity of the boot process and the integration of various drivers and functionalities increase the attack surface, providing adversaries with numerous opportunities to exploit vulnerabilities~\cite{eclypsium_spi_write_protections}.

Recent years have witnessed a sharp increase in UEFI-related attacks \new{reflected in the growing number of UEFI-specific entries in CISA's Known Exploited Vulnerabilities catalog}~\cite{eclypsium2024kev}, with notable examples discussed in the following chapters.
The challenge of detecting and removing threats at the UEFI level highlights the need for specialized security solutions and practices designed to protect against, and respond to, such deep-seated attacks.

\subsection{Scope and Outline}
We begin by exploring UEFI's functionality and inherent security features in Section~\ref{sec:UEFI-introduction} and then study the UEFI-related attacks found in the wild such as BlackLotus, MoonBounce, FinSpy, etc. in Section~\ref{sec:uefi_attacks}. Following this study of past attacks, in Section~\ref{sec:uefi-mitre} we present a MITRE-like~\cite{mitre} taxonomy that includes the tactics, techniques, and sub-techniques employed to carry out UEFI-related attacks. 
In Section~\ref{sec:uefi-countermeasures} we create a taxonomy of countermeasures that can be used to protect the system against UEFI-related attacks.
We conclude the paper by performing a risk analysis of the UEFI security landscape in Section~\ref{sec:riskanalysis} to find out the most vulnerable components of the UEFI ecosystem.
\subsection{Related Work}
\label{subsec:Related_works}
The early investigation into the vulnerabilities within UEFI firmware by Bashun et al. in "Too Young to Be Secure: Analysis of UEFI Threats and Vulnerabilities" (2013)~\cite{RelWorkinproceedingsTooyoung} set the stage for subsequent research in this relatively uncharted field.
Building on the foundational concerns, Rodionov et al.'s "Bootkits: Past, Present \& Future" (2014)~\cite{RelWorkrodionov2014bootkits} traced the evolution of rootkits, including those targeting UEFI. This survey provided an understanding of how UEFI threats have progressed over time, and indicated a trend toward more sophisticated and targeted attacks.
Diving deeper into specific threats, Wang and Dong's "Attacking Intel UEFI by Using Cache Poisoning" (2019)~\cite{CAR} gave details about a novel attack method, illustrating the advanced nature of threats to UEFI systems. This specificity in research pinpointed vulnerabilities that could be exploited by attackers with a high degree of technical knowledge.
With the threat landscape evolving, recent studies have focused on proactive security enhancements. Krichanov and Cheptsov's work on "UEFI Virtual Machine Firmware Hardening Through Snapshots and Attack Surface Reduction" (2021)~\cite{RelWorkkrichanov2021uefidet_mitigation} introduced a methodological approach to defend against UEFI threats in virtualized environments. This work demonstrated the potential of leveraging virtualization technology to safeguard UEFI firmware.
In a more recent setting, Jiao et al.'s "UEFI Security Threats Introduced by S3 and Mitigation Measures" (2022)~\cite{RelWorkjiao2022uefi} targeted the vulnerabilities present in the S3 sleep state, offering mitigation strategies tailored to the nuances of power state transitions in UEFI. This indicates a growing awareness of the need to secure all aspects of UEFI functionality.
The most recent survey by Zhou et al., "A Survey on the Evolution of Bootkits Attack and Defense Techniques" (2024)~\cite{RelWorkzhou2024survey}, along with a practical exploration of rootkit design in "The Design of the Simple SMM Rootkit" (2022)~\cite{SMMrootkit}, collectively indicate a sustained interest in UEFI security. These works contribute to a broader understanding of how attackers may exploit UEFI, and the defensive techniques that could be employed in response.

Despite the progress made in identifying and mitigating UEFI security threats, the field remains under-researched.
The existing literature, while providing isolated insights, falls short of establishing a comprehensive taxonomy of UEFI-related attacks. 
Such a taxonomy is essential for several reasons---it serves as a foundation for recognizing common patterns in attacks which could be an essential building block for organizations' Security Information and Event Management (SIEM) and Threat Hunting systems; facilitates the systematic arrangement of countermeasures; and enables the identification of the most exploitable attack vectors.
Hence, there is a pressing need for a robust taxonomy of UEFI-related attacks. The lack of such a taxonomy hinders the ability to conduct a thorough risk analysis, which is critical for determining the most exploitable weaknesses within UEFI systems.
The countermeasures against UEFI threats are often discussed in isolation, without the context of an overarching defense strategy. 
A structured presentation of these defenses, correlated with a detailed taxonomy of attack techniques, is essential for a holistic understanding of UEFI security and the efficacy of various protection mechanisms.
This research addresses a gap in the existing research on UEFI security by presenting a comprehensive framework. 
The framework encompasses a taxonomy of UEFI threats, a systematic arrangement of countermeasures, and a risk analysis that identifies and prioritizes the most critical vulnerabilities. 
The objective is to guide future research and development endeavors, focusing on fortifying the most vulnerable aspects of UEFI security.

\subsection{Contributions}
\label{subsec:contributions}

The main contributions of this paper can be summarized as follows: 
\begin{enumerate}[nosep]
    \item \emph{Formalization of UEFI Security Knowledge:} Systematically formalizing existing UEFI security knowledge by examining its history, development lifecycle, supply chain, and functionalities.
    \item \emph{Taxonomy of UEFI-Related Attacks:} Developing a UEFI-specific taxonomy of attack tactics and techniques, mapping the attack landscape based on existing literature, and presenting it in a MITRE-like layout for ease of use.
    \item \emph{Taxonomy of UEFI-Related Security Controls:} Enumerating and categorizing security controls designed to defend against UEFI-related attacks, and providing a reference guide for potential threats.
     \item \emph{Comprehensive Risk Analysis:} Performing a thorough risk analysis that identifies defense gaps and suggests urgent research directions to enhance UEFI security.
\end{enumerate}

\section{Background}
\label{sec:UEFI-introduction}
Since the inception of computer systems, the Basic Input/Output System (BIOS) has been essential for hardware initialization during the boot process~\cite{thompson2003BIOS}.
As the first code that runs in a computer system, its primary tasks are to initialize the hardware and launch the OS. 
However, with advancements in computing, several limitations of legacy BIOS have become apparent. 
These include the inability to boot from hard drives larger than 2.2 TB, slow boot times, the absence of inherent security mechanisms, a fairly simplistic user interface, and a lack of extensibility due to its monolithic design and CPU-dependent architecture~\cite{RelWorkinteljounel15, BIOSlimitations}. 

To address these shortcomings, the UEFI Forum~\cite{uefi_member}---an industry-wide group comprising hardware manufacturers, major chip manufacturers, motherboard manufacturers, and OS vendors---released the Unified Extensible Firmware Interface (UEFI) specifications in 2005, aimed at fostering the development of a successor to legacy BIOS. 

\subsection{Components Involved in UEFI Boot}

The following components are involved in the UEFI boot process~\cite{uefi_PI_1_8}:

    \subsubsection{SPI Flash Chip:} The Serial Peripheral Interface (SPI) flash chip, often referred to as the BIOS or firmware flash, is a non-volatile storage medium residing on the motherboard. 
    It stores firmware that initializes hardware and provides services for the system's software~\cite{wilkins2013uefi}.
    The exact contents of the SPI chip vary depending on system manufacturers, motherboard or device models, UEFI firmware vendors, and system configuration: 
    \begin{itemize}
        \item \emph{UEFI Firmware:} This is the primary content of most SPI flash chips on PCs.
        It initializes hardware, provides low-level services, and then hands off control to the operating system.
        UEFI uses the physical flash device as storage.
        UEFI firmware consists of firmware volumes, which contain firmware files, and these files contain the most basic unit---the firmware segment.
        
        \item \emph{UEFI Variables and Configuration Data:} On some systems, UEFI variables (such as boot order, security keys for Secure Boot, and other configuration settings) are stored in a dedicated region of the SPI flash chip.
        Not all UEFI variables are hard-coded onto the SPI chip. 
        Many UEFI variables, such as boot order and system settings, are dynamically set and modified during system boot or other run-operations.
        Some variables, especially those related to system security, like Secure Boot keys, 
        might be provisioned during manufacturing and are not intended to be changed under normal circumstances~\cite{UEFIvarPkKekDbDbx}.
        A few common such variables are:
        \begin{enumerate}
            \item \emph{Platform Key (PK):} Used in the Secure Boot process to verify the Key Exchange Key (KEK) and control its updates.
            \item \emph{Key Exchange Key:} Used to verify db and dbx updates. 
            \item \emph{Allowed Signatures Database (db):} Contains keys and signatures of allowed bootloaders and drivers.
            \item \emph{Forbidden Signatures Database (dbx):} Contains revoked certificates of forbidden bootloaders and drivers in the form of their hashes.
        \end{enumerate}
        
        \item \textit{System Management Mode (SMM) Code:} SMM is a special operating mode in x86 microprocessors introduced by Intel as a security and system-management feature.
        It provides an isolated environment for system-level operations that are transparent to the operating system, applications, and other components.
        It handles system-wide functions like power management, hardware control, or proprietary OEM-designed code.
        Code executed in this mode can be stored on the SPI flash chip~\cite{SMMrootkit}.

        \item \emph{Option ROM:} These firmware pieces help initialize specific hardware, such as graphics cards or certain storage controllers. 
        
        \item \emph{Recovery Firmware:} Some systems include a backup or recovery version to restore the firmware if the primary copy becomes corrupted~\cite{RelWorknystrom2011uefinetworking}.
        
        \item \emph{Certificates and Keys:} For systems that support Secure Boot or other security features, necessary certificates and keys might be stored in the SPI flash~\cite{RelWorknystrom2011uefinetworking}.
        
        \item \emph{Diagnostic Tools or Utilities:} \new{Some manufacturers include built-in diagnostic tools or utilities in the firmware, accessible during boot~\cite{RelWorkinteljounel15}. Although these components are primarily intended for maintenance and troubleshooting, if they are exposed insecurely or insufficiently protected, they may expand the pre-OS attack surface and provide additional functionality that could be abused by an attacker.}
    \end{itemize}  
    
    \subsubsection{Serial Peripheral Interface (SPI):} It is a synchronous serial communication interface used for short-distance communication, primarily in embedded systems~\cite{uefiSPI}.
    The SPI flash chip communicates with the CPU using the SPI protocol over an SPI Bus.
    When the system powers on, the CPU communicates with the SPI flash chip through the SPI controller to fetch the initial firmware code for execution.
    This bootstrapping process begins when the CPU starts the system's firmware, which initializes the rest of the system and eventually loads the operating system~\cite{wootton2016serial}.
    
    \subsubsection{EFI System Partition (ESP):}
    The ESP is a crucial component of the UEFI firmware architecture.
    Its primary purpose is to hold the bootloader and related files necessary for the computer to boot.
    The ESP is a dedicated partition on a storage device, typically a hard drive or solid-state device (SSD), that stores essential boot-related files and is created during the installation of the operating system.
    The installer typically creates a small, dedicated partition at the beginning of the storage device.
    It formats this partition with the FAT32 file system and assigns the EFI System Partition label to it.
    Once the ESP is created, the installer copies the necessary bootloader and related files to it.
    The installer often configures the UEFI bootloader to point to the location of the OS files on the storage device.
    This configuration is stored in the boot configuration data (BCD) within the ESP~\cite{matrosov2019rootkits}.
    On machines with multiple operating systems, each OS typically stores its bootloader in the ESP.
    The UEFI firmware then presents a boot menu, allowing the user to choose which OS to start.

    If Secure Boot is enabled, the installer may also generate and install Secure Boot keys and certificates.
    These keys verify the authenticity of the bootloader and other crucial components during the boot process~\cite{RelWorkhagl2021secureboot}.

\subsection{UEFI Boot Process}
\label{appx:uefi_booting_process}
The UEFI standard specifies a flexible environment for the pre-boot and boot processes in computer systems. 
It outlines the essential features and requirements for UEFI firmware, covering crucial aspects like booting, hardware initialization, and run-time services~\cite{uefi_specs_2_10}.
The UEFI PI (platform initialization) is a subset of the UEFI specification that defines a standardized set of boot and run-time services for platform initialization~\cite{uefi_PI_1_8}.
While the services and their interfaces are standardized, UEFI implementations can differ across computer manufacturers. 
Each manufacturer typically develops its own UEFI firmware based on a reference implementation~\cite{SBOM-supplychain}.

From a security perspective, the boot process consists of two distinct parts. 
The first is the platform initialization process, independent of the OS installed on the system~\cite{uefi_PI_1_8}. It starts when the system powers on and continues until the execution of DXE phase is completed, after which control is handed from UEFI to the target OS~\cite{zimmer2017beyond}.
The second part is specific to the OS being chosen for boot, differing for Windows, Mac, Linux, etc.~\cite{matrosov2019rootkits,uefi_specs_2_10}

According to the UEFI specification, the entire system boot process consists of six phases, as depicted in Figure~\ref{fig:UEFI_BootManager_flow}, during which different tasks are performed. 
Of these, the following phases are OS-independent:
\begin{enumerate}
\item 
\noindent \textbf{SEC Phase:} 
        The SEC phase begins immediately after the system powers on or resets. 
        It starts by running UEFI code from the SPI chip on the motherboard.
        During this phase, the CPU and chipset are initialized, and system memory is set up. 
        The SEC phase also includes configuring of the CPU cache as RAM (CAR), establishing a temporary memory store. 
        CAR then facilitates the execution of subsequent UEFI firmware initialization stages~\cite{uefi_PI_1_8}. 
\item 
\noindent\textbf{Pre-EFI Initialization (PEI) Phase:}
         Various PEI modules (PEIMs) are executed during the PEI phase to handle specific initialization tasks.
         These modules include platform-specific initialization code, memory initialization routines, and CPU initialization. 
         The last PEIM executed during this phase is DXE IPL (Initial Program Load), which handles the transition to the DXE phase~\cite{uefi_PI_1_8}. 
\item         
\noindent \textbf{Driver Execution Environment (DXE) Phase:}
        Most of the system initialization takes place in the DXE phase. 
        During this phase, device drivers are loaded.
         It also creates tables for EFI boot and run-time services. 
         
\item 
 \noindent\textbf{Boot Device Selection (BDS) Phase}: In this phase, the UEFI boot manager hands control to the OS boot manager selected by the user (e.g., \emph{bootmgfw.efi} for Windows).
 After selecting the OS boot manager, boot behavior is determined by settings stored in the Boot Configuration Data (BCD) file.
 The boot manager then loads the boot loader, such as \emph{winload.efi} for Windows.
 For execution, \textit{winload.efi} also relies on the information in the BCD. 
 All of these files (\textit{bootmgfw.efi, BCD, winload.efi}) reside in the EFI System Partition (ESP) of the hard drive.
 
 \begin{figure}[h]
 \caption{UEFI booting process}
    \centering    \includegraphics[width=0.99\textwidth]{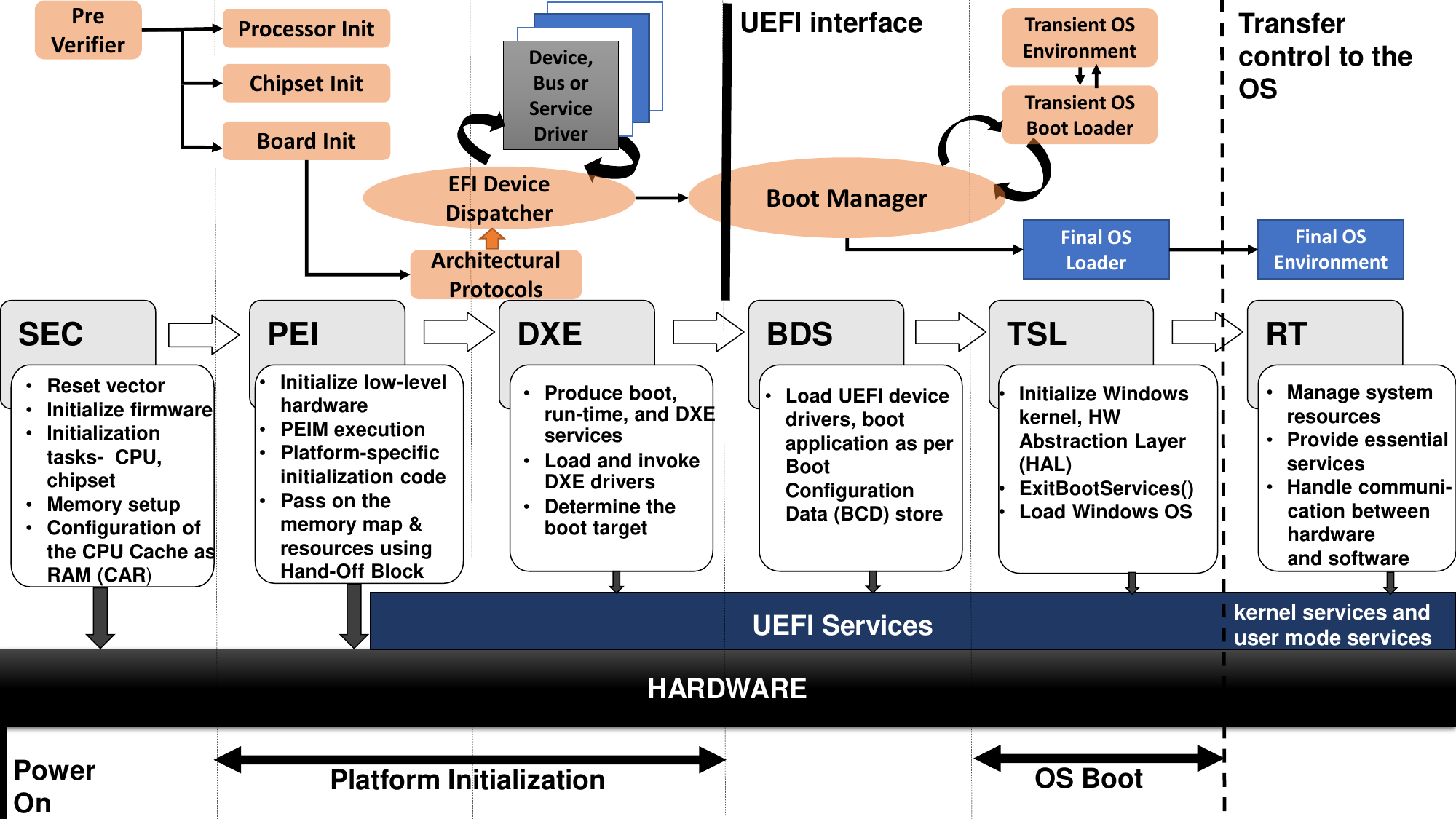}    
    
    \label{fig:UEFI_BootManager_flow}
\end{figure}
\item 
\noindent\textbf{TSL Phase:} OS kernel initialization takes place in this phase.
When initialization is complete, the \textit{ExitBootServices()} function (from UEFI boot services) is called, which marks the completion of the boot process. 
After executing \textit{ExitBootServices()}, only UEFI run-time services remain available, while UEFI boot-time services are no longer accessible~\cite{uefi_specs_2_10}. 
\item 
\noindent\textbf{Run-Time:} After the kernel initialization is complete in the TSL phase, control is handed over to the OS run-time phase~\cite{matrosov2019rootkits}.
\end{enumerate}
\subsection{Security Features of UEFI}

UEFI provides several security features to protect the integrity of the boot process and OS components.
These features aim to enhance the security of the boot process, firmware, and overall system integrity.
This section explores some key security features available in UEFI.

    \subsubsection{Secure Boot:} The primary objective of Secure Boot is to ensure that only trusted software loads during the boot process, preventing the execution of unauthorized or malicious code~\cite{eclypsiumsecureboot}.
    Secure Boot verifies digital signatures to validate the integrity and authenticity of the boot components.
    This mechanism safeguards against malware that may attempt to exploit the boot process~\cite{matrosov2019rootkits}. 
    Microsoft bootloaders, being Secure Boot-aware, leverage UEFI Secure Boot keys and databases.
    During the boot process, these bootloaders verify the digital signatures of firmware, drivers, and OS components against keys stored in the UEFI firmware.
    This verification ensures that only trusted and signed components load, preventing the execution of unauthorized or malicious~code.
  
    Linux bootloaders follow a different approach to key management for Secure Boot. 
    They rely on the machine owner key (MOK) and shim to switch to a key chain provided by Red Hat or Canonical, rather than using the UEFI variables directly.
    This allows Linux distributions to use their own set of keys and enable Secure Boot functionality without relying solely on UEFI firmware keys~\cite{mok_secboot}.
   This approach allows Linux distributions to control the Secure Boot process and provide a Secure Boot environment for users.
    
    Both approaches---Microsoft bootloaders leveraging UEFI Secure Boot keys and Linux bootloaders using MOK and Shim---aim to verify the integrity and authenticity of boot components.

    \subsubsection{Secure UEFI Variable Storage:} UEFI incorporates a robust storage mechanism to secure variables.
    These variables are safeguarded against unauthorized modifications by external parties~\cite{accesscontrol}.

    \subsubsection{Platform Key and Signature Database:} UEFI firmware provides native support for cryptographic keys, including PK and KEK.
    These keys play a vital role in verifying the integrity and authenticity of boot components~\cite{wilkins2013uefi}.
    The PK holds the highest significance, serving as a root of trust within the system.
    KEKs facilitate the management of other keys within the UEFI environment~\cite{UEFIvarPkKekDbDbx}.
    UEFI firmware also maintains a signature database that houses certificates and hashes of trusted boot components~\cite{wilkins2013uefi}.
    These databases serve as references for validating the authenticity and integrity of boot loaders, drivers, and firmware updates.
    
    \subsubsection{Credential Storage:} UEFI firmware can securely store credentials, such as Secure Boot keys or BitLocker recovery keys, to safeguard against unauthorized access.
    By securely storing these credentials, UEFI firmware ensures that only authorized entities can retrieve and use them, preventing unauthorized access to critical system components or data~\cite{bitlocker}. 
    
    \subsubsection{Trusted Platform Module (TPM) Integration:} The TPM is a dedicated cryptographic hardware module that enhances system security.
    When integrated with UEFI, TPM offers security capabilities, including secure key storage, Secure Boot measurements, and remote attestation~\cite{uefi_TCG}.
    
    \subsubsection{Measured Boot:}Measured Boot is a UEFI feature that captures a cryptographic hash, or measurement, of every component loaded during the boot process~\cite{RelWorknystrom2011uefinetworking}.
    These measurements can be securely stored in the TPM and used for attestation purposes, thereby ensuring the security of the boot process.
    During each boot, Measured Boot stores the cryptographic hash of each loaded component, including boot loaders, firmware, and OS files.
    These measurements are securely stored within the TPM, providing a tamper-resistant environment for their retention.
    The stored measurements can be accessed later for attestation, which enables a trusted entity to remotely verify the integrity and security of the boot process.
    By comparing stored measurements with expected values, the system can provide evidence of a Secure Boot process, guarding against tampering or unauthorized modifications~\cite{nsa_uefi_defensive_practices}.
        
    \subsubsection{UEFI Capsule Updates:} 
     The host OS generally provides support for updating the underlying system firmware.
    This makes the protection of such updates a critical component of the overall system security. 
    The firmware update is passed from the host OS to UEFI in a capsule, which uses digital signatures to validate the integrity and authenticity of firmware update packages~\cite{SMMrootkit}.
    This mechanism establishes a secure framework for delivering firmware updates while safeguarding against unauthorized or malicious firmware installations.
    
    \subsubsection{BIOS Write Protection:} Certain UEFI firmware implementations incorporate BIOS write protection. If enabled, it prevents unauthorized modifications to the firmware on the chip~\cite{RelWorkkallenberg2014defeatingsecboot}.

    \subsubsection{User Authentication:} UEFI incorporates a user authentication mechanism to restrict access to and modification of UEFI settings to authorized individuals~\cite{UEFIvarPkKekDbDbx}.
    
    \subsubsection{Hardware-Based Trusted Execution Environment (TEE):} Modern CPUs increasingly incorporate built-in TEEs, secure areas of the CPU that ensure confidentiality and integrity of the code and data loaded within it.
    Leveraging TEEs, UEFI can establish a secure execution environment that isolates sensitive operations and data from potential threats.
    TEEs provide hardware-based security features, including secure memory enclaves and encrypted virtualization, which UEFI can harness to enhance its security capabilities.
    Leveraging TEEs can ensure the confidentiality and integrity of critical system components, protect against memory-based attacks, and strengthen the overall security posture of the system~\cite{zimmer2016establishing}.

\section{UEFI Attacks}
\label{sec:uefi_attacks}
\subsection{Attacks Observed in the Wild}
\begin{figure*}[ht]
 \caption{UEFI Attack Timeline}
    \centering
    \includegraphics[width=1\textwidth]{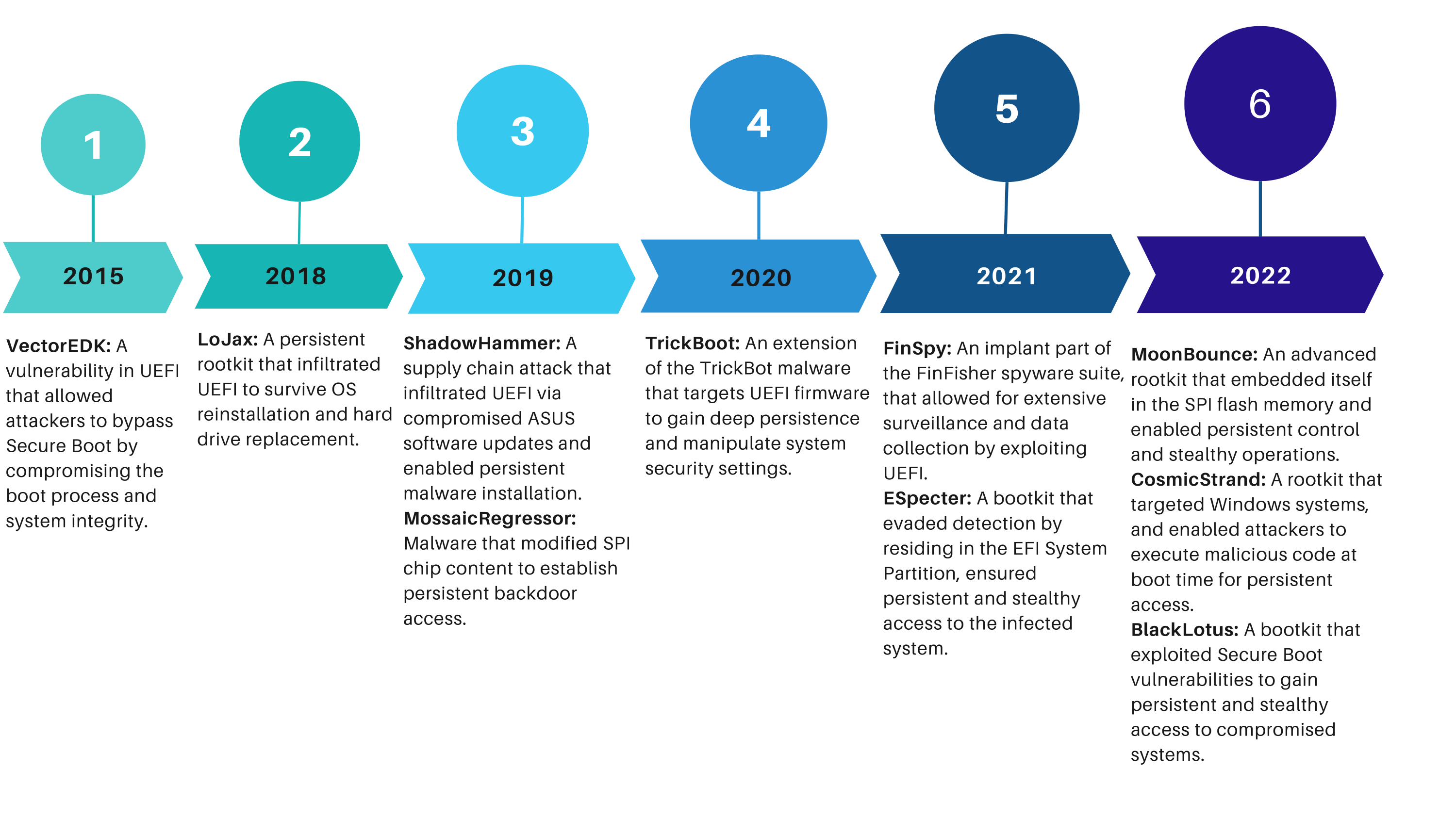}    
    
    \label{fig:attack_timeline}
\end{figure*}

To identify UEFI-specific threats and vulnerabilities exploited in attacks, we surveyed the academic literature and publicly available technical security reports. We discuss these in the sections below, provide a timeline in \autoref{fig:attack_timeline}, and compare them in Table~\ref{tab:attacks_table}.

From reviewing the attacks, we can draw a few conclusions and identify several trends. 
\begin{enumerate}
    \item \textit{Evolution of Threat Actors:} In the early phase, (2009-2013), attacks were conducted by well-resourced state actors aiming to carry out covert espionage activities. For example, DarkSeaSkies~\cite{DarkSeaSkies} and DerStrake~\cite{DerStarke2} were tools developed by the Central Intelligence Agency (CIA) for manipulating UEFI. SonicScrewDriver~\cite{SonicScrewDriver} was another CIA tool that enabled attackers to infect UEFI through USB ports.
    After 2013, the UEFI attack landscape diversified significantly. During this period, the types of threat actors expanded geographically, with a shift in motivations. Financial gain became a significant driver alongside traditional state-sponsored espionage.
    Recently, UEFI attack tools that were once exclusively available to highly sophisticated actors have become increasingly accessible to a broader audience. These tools are now for sale on various online platforms, such as hacking forums and dark web marketplaces.
    \item \textit{Increasing Complexity of Attacks:} Modern UEFI malware use more advanced stealth and persistence techniques, as seen in BlackLotus~\cite{blacklotus1} and MoonBounce~\cite{moonbounce1}, where attackers employed sophisticated hooking mechanisms to conceal their presence and move to memory locations where detection is extremely difficult. Attackers also employed methods such as modifying UEFI variables, disabling HVCI and PatchGuard~\cite{blacklotus1}, and bypassing kernel protection and driver signature enforcement~\cite{Dreamboot1,especter1}.
    \item \textit{Shift in Attack Objectives:} A shift in objectives is evident, with attacks moving from being purpose-specific to becoming purpose-agnostic. Initial UEFI attacks focused on specific disruptions, whereas recent attacks are directed towards long-term espionage~\cite{mosaicregressor1}, surveillance~\cite{finspy1}, targeted attacks~\cite{shadowhammer1} and data exfiltration~\cite{cosmicstrand1}.
    \item \textit{Increased Frequency:} The number of detected attacks has risen in recent years.
\end{enumerate}

\subsection{Attack PoCs}
It is important to note that besides the attacks discovered in the wild, the research community has released proof-of-concepts for additional attacks, many of which are available on public code repositories. For completeness, we address these in the attacks Table \ref{tab:attacks_table} as well.

\begin{landscape}
\begin{tiny}
\setlength{\tabcolsep}{3pt}
\newcolumntype{L}[1]{>{\raggedright\arraybackslash}p{#1}}
\begin{longtable}{@{}L{1cm}L{0.4cm}L{0.6cm}L{2cm}L{1.8cm}L{2cm}L{1.8cm}L{2cm}L{2cm}L{2cm}L{2cm}L{0.8cm}@{}}
\caption{UEFI-related Attacks}
\label{tab:attacks_table}\\
\toprule
{Name} & 
{Year} & 
{PoC/ Attack} & 
{Description} & 
{Components} &
{Initial Access} & 
{Initialization and }{Privilege Escalation} & 
{Establish} {Foothold} & 
{Maintain} {Persistence} & 
{Post-Exploitation} & 
{Actions} {on Objective} & 
{Ref.} \\
\midrule
\endhead
Dark- SeaSkies & 2009 & Attack & Malicious implant for Apple MacBook Air. & DarkMatter, SeaPea, and NightSkies. & Insufficient public data available. & & & & & &\cite{DarkSeaSkies} \\
\midrule
Sonic Screwdriver & 2012 & Attack & CIA tool. Exploits Thunderbolt interface. Bypasses firmware password via DMA. & Insufficient public data available. & Requires physical access. Uses a modified Thunderbolt-to-Ethernet adapter. & Overrides boot process security measures. & Embeds within device firmware. & Resides in EFI partition. Survives system reboots. & Surveillance. Data extraction. & Installs additional espionage tools. Manipulates device operations. & \cite{SonicScrewDriver}\\
\midrule
DerStrake & 2013 & Attack & Advanced automated implant for macOS, operates without relying on a physical disk. & Insufficient public data available. & Insufficient public data available.& Maintains presence within the computer's EFI firmware. & Operates within the disk arbitration process of macOS. & Highly resistant to detection and removal due to EFI firmware embedding. & Often uses network communications through web browsers to evade detection. & Designed for stealthy operations, potentially enabling espionage or data extraction. & \cite{DerStarke1, DerStarke2, DerStarke3}\\
\midrule
DreamBoot & 2013 & PoC & Targets UEFI to attack OS bootloader, requires Secure Boot disabled.& Hooking functions, malicious payload. & Loaded via modified ISO with EFI PE binary. & Modifies bootmgfw.efi and winload.efi, bypasses kernel protections. & Integrates into bootloader and kernel loader. & Embedded at firmware level. & Disables kernel protections. Executes arbitrary code. & Corrupts Windows kernel. Bypasses authentication. & \cite{Dreamboot1,Dreamboot2}\\
\midrule
Thunder- strike & 2014 & PoC & Targets macOS via Thunderbolt, implants firmware-level malware.& Hooking functions, malicious payload. & Requires physical access. Uses a malicious Thunderbolt device. & Exploits firmware via Unsigned Option ROMs. & Modifies boot ROM to embed malicious code. & Resides within UEFI firmware. & Can launch further attacks. Deploy malware. & Aims for undetectable, persistent root-level access. & \cite{thunderstrike1,thunderstrike2}\\
\midrule
DarkJedi / Thunderstrike2 & 2014 & PoC & Modifies PEI core firmware, targets UEFI boot script table. & Hooking functions, malicious payload. & Through malicious Thunderbolt devices. & Writes to boot flash during PEI phase. & Infects motherboard firmware. & Embeds in firmware. Survives reinstalls/reboots. & Can propagate via infected devices. & Controls firmware/hardware. Leads to data theft. & \cite{darkjedi}\\
\midrule
VectorEDK & 2015 & Attack & Part of Hacking Team leak, embeds malware in UEFI. & NTFS parser (Ntfs.efi), rkloader.efi, fsbg.efi, scout.exe, soldier.exe. & Likely through physical means or USB devices. & Installs DXE driver pre-OS. & Embeds in UEFI, includes NTFS interaction modules. & Resides in UEFI, survives OS reinstalls. & Can disable security features, spy on users. & Espionage. Data theft. System manipulation. & \cite{vectoredk1,vectoredk2} \\
\midrule
LightEater & 2015 & PoC & Targets SMM to access physical memory, seeks sensitive data. & Insufficient public data available.
 & Needs physical access or exploit. & Inserts/modifies SMM handlers. & Embeds in SMM. & Remains active across reboots/OS reinstalls. & Targets encryption keys, and sensitive data. & Data harvesting, potential malware installation. & \cite{lighteater}\\
\midrule
PeiBackdoor & 2015 & PoC & Enables arbitrary code execution during PEI phase.& Insufficient public data available. & Requires access for firmware modification. & Hooks into early UEFI code. & Integrates within PEI phase. & Active from early boot stages. & Alters boot process. Modifies firmware settings. & System surveillance. Further malware deployment. & \cite{peibackdoor} \\
\midrule
Thinkpwn & 2016 & PoC & Exploits UEFI vulnerability in Lenovo ThinkPad models.& Insufficient public data available. & Requires admin privileges or physical access. & Modifies UEFI firmware directly. & Embeds in firmware, loads before OS. & Persists through restarts and OS changes. & Manipulates system functions. Disables Secure Boot. & Espionage. System disruption. Backdoor creation. & \cite{thinkpwn} \\
\midrule

LoJax & 2018 & Attack & Exploits LoJack anti-theft software for UEFI-level persistence. & SecDxe DXE driver, NTFS driver, malicious payload (Autoche.exe), small agent (Rpcnetp.exe). & Spear-phishing or compromised websites. & Modifies UEFI firmware after gaining admin rights. & Embeds within UEFI firmware. & Survives OS reinstalls and disk replacements. & Executes additional malicious payloads at boot. & Long-term espionage. Data exfiltration. & \cite{lojax1,lojax2,lojax3} \\
\midrule
Mosaic- Regressor & 2019 & Attack & Modular cyber-espionage framework targeting diplomatic entities. & NTFS driver, SMM reset UEFI application, SmmInterfaceBase bootkit, SmmAccessSub persistent dropper. & Likely via phishing or physical access to deploy compromised firmware. & Embeds within UEFI firmware at boot. & Executes with each system boot. & Remains after hard drive changes/OS reinstalls. & Downloads additional modules. Performs espionage activities. & Espionage. Network compromise. Data theft. & \cite{mosaicregressor1,mosaicregressor2,mosaicregressor3}\\
\midrule
Shadow- Hammer & 2019 & Attack & Targeted ASUS Live Update Utility, affecting supply chain. & Insufficient public data available. & Compromised ASUS utility distributed via official channels. & Executes a backdoor payload upon update. & Uses legitimate certificates to avoid detection. & Minimal changes to software's file size/signature. & Executes further malicious payloads. & Targeted espionage based on predefined MAC addresses. & \cite{shadowhammer1,shadowhammer2,shadowhammer3,shadowhammer4}\\
\midrule
Trickboot & 2020 & Attack & TrickBot module that probes and exploits UEFI/BIOS vulnerabilities. & PermaDll Module- reconnaissance tool, Firmware Interaction Functions. & Spear-phishing or web vulnerabilities. & Checks and exploits UEFI/BIOS settings. & Writes payloads directly into UEFI firmware. & Operates below OS, survives system reinstalls. & Disables security measures, downloads payloads. & Espionage. Malware delivery. Network compromise. & \cite{trickboot1,trickboot2,trickboot3,trickboot4} \\
\midrule
BootHole & 2020 & PoC & Vulnerability in GRUB2 affecting Secure Boot, allows arbitrary code execution. & Insufficient public data available. & Requires admin privileges to modify GRUB2 configuration. & Exploits buffer overflow in GRUB2. & Manipulates boot process via GRUB2 config. & Manipulates bootloader for consistent execution. & Installs malware. Modifies OS loading. & Alters OS loading. Executes arbitrary actions at startup. & \cite{boothole1}\\
\midrule
ESPecter & 2021 & Attack & Bypasses Windows driver signature enforcement, persists in ESP. & Compromised Boot Manager, malicious payloads & Likely through exploiting vulnerabilities or social engineering. & Modifies Windows Boot Manager early in the boot process. & Patches system functions to bypass integrity checks. & Modifies ESP, targeting boot-related files. & Deploys kernel-mode driver, sets keyloggers, communicates with C\&C. & Executes commands. Downloads malware. Controls compromised machine. & \cite{especter1,especter2,especter3}\\
\midrule
FinFisher Finspy & 2021 & Attack & Surveillance tool that can infect via trojanized installer or UEFI bootloader. & Compromised Boot Manager, Encrypted Payloads. & Exploits system vulnerabilities or physical access. & Embedded at firmware level, operates before OS boots. & Replaces Windows Boot Manager with a malicious one. & Resides in UEFI firmware, unaffected by system changes. & Decrypts/executes payloads, and establishes backdoor. & Surveillance. Data exfiltration. Further malware delivery. & \cite{finspy1,finspy2,finspy3}\\
\midrule
Moon- Bounce & 2022 & Attack & Targets SPI flash chip's UEFI firmware, modifies CORE\_DXE component. & Function hooks, malicious payload. & Through firmware vulnerabilities or supply chain compromises. & Manipulates early boot process, controls system pre-OS load. & Resides in non-volatile SPI flash memory. & Embedded in firmware, unaffected by disk/OS changes. & Manipulates boot sequence, loads malicious components. & Network reconnaissance. Data exfiltration. Deploying further payloads. & \cite{moonbounce1,moonbounce2}\\
\midrule
Cosmic- Strand & 2022 & Attack & Infects UEFI of specific motherboards, used for cyberespionage.& Function hooks, malicious payload. & Speculated physical access or supply chain compromise. & Modifies CSMCORE DXE driver, patches entry to redirect. & Sets up hooks in Windows boot process. & Alters UEFI, affects even after OS reinstall/drive replacement. & Executes malicious component inside Windows. & Targeted data collection. Surveillance. Persistent presence. & \cite{cosmicstrand1,cosmicstrand2,cosmicstrand3,cosmicstrand4} \\
\midrule
Black- Lotus & 2022 & Attack & First UEFI attack to bypass Secure Boot, sold on the dark web. & Malicious bootloader, malicious BCD, legitimate but vulnerable binary executables. & Installs via legitimate but vulnerable Windows binaries, possibly online/offline. & Disables Secure Boot using CVE-2022-21894. & Sets up legitimate shim binary as default bootloader. & Deploys kernel driver to prevent removal. & Deploys user-mode components, executes kernel payloads. & Full control over OS boot process. Deploys arbitrary payloads. & \cite{blacklotus1} \\
\bottomrule

\end{longtable}
\end{tiny}
\end{landscape}

\section{UEFI MITRE ATT\&CK-like Mapping}
\label{sec:uefi-mitre}
During the investigation of UEFI attacks (as summarized in Table~\ref{tab:attacks_table},  we concluded that the various attack patterns can be broken down into basic elements and documented in a similar way to other domains.
Consequently, we created a MITRE-like mapping of the tactics and techniques used in UEFI-related attacks. This mapping can be applied to both old and new attacks, categorizing them by TTPs (Tactics, Techniques, and Procedures) and stages, providing a common language for security professionals and researchers. We acknowledge that this mapping may not be exhaustive, and we invite the UEFI security research community to adopt it and expand it. In creating the framework, we intentionally excluded non-UEFI perspectives, even if they are relevant to specific UEFI attack campaigns. For example, network scanning might be part of a UEFI-related campaign; however, since it is not specific to UEFI, it is not included in the framework. This decision is based on the understanding that cybersecurity attacks are often complex, involving a variety of tools, techniques, and capabilities that may extend beyond a specific sub-domain. Therefore, we focused on UEFI-specific components of the attack, which we outline here, while excluding non-UEFI-specific elements that are covered in existing literature and other frameworks. 
To demonstrate the practical utility of such a mapping, we use it as a basis for conducting a comprehensive risk analysis of UEFI-related security concerns in \cite{mitre} where we map existing security countermeasures against each technique.
A brief description of the attack techniques follows. For a detailed description of each technique, we encourage readers to refer to the respective references in Section~\ref{Tab:UEFI_TTPs}. 
\clearpage
\begin{tiny}
\begin{longtable}{>{\raggedright\arraybackslash}p{2cm} 
                  >{\raggedright\arraybackslash}p{4cm} 
                  >{\raggedright\arraybackslash}p{4cm} 
                  >{\raggedright\arraybackslash}p{2cm}}
\caption{UEFI Techniques and Sub-Techniques}
\label{Tab:UEFI_TTPs}\\
\toprule
\textbf{Tactic} & \textbf{Technique} & \textbf{Sub-Technique} & \textbf{Reference} \\
\midrule
\endfirsthead
\toprule
\textbf{Tactic} & \textbf{Technique} & \textbf{Sub-Technique} & \textbf{Reference} \\
\midrule
\endhead
\midrule
\endfoot
\bottomrule
\endlastfoot

Reconnaissance (UEFI-TA9001) 
  & System Enumeration (UEFI-T0001) 
  & - 
  & \cite{blacklotus1}, \cite{trickboot2}, \cite{trickboot3}, \cite{shadowhammer1} \\

  & UEFI Code Analysis and Vulnerability Research (UEFI-T0002) 
  & - 
  & \cite{trickboot1}, \cite{lojax1}, \cite{blacklotus1} \\

  & Hardware and UEFI Inventory (UEFI-T0003) 
  & - 
  & \cite{trickboot1}, \cite{trickboot3} \\

  & Supply Chain Identification (UEFI-T0004) 
  & UEFI Development Process (UEFI-T0004.001) 
  & \cite{shadowhammer1} \\

  & 
  & Distribution Chain (UEFI-T0004.002) 
  & \cite{shadowhammer3}, \cite{shadowhammer4} \\

  & 
  & Third-Party Dependency (UEFI-T0004.003) 
  & \cite{shadowhammer1}, \cite{SBOM-supplychain} \\

\midrule

Resource Development (UEFI-TA9002) 
  & Acquire Access (UEFI-T0005) 
  & 
  & \cite{trickboot1}, \cite{shadowhammer1} \\

  & Acquire Infrastructure (UEFI-T0006) 
  & 
  & \cite{mosaicregressor2}, \cite{lojax3} \\

  & Compromise Accounts (UEFI-T0007) 
  & 
  & \cite{shadowhammer1} \\

  & Develop Capabilities (UEFI-T0008) 
  & Exploit Development (UEFI-T0008.001) 
  & \cite{paloalto_glupteba_malware}, \cite{lojax3} \\

  & 
  & Certificate Management (UEFI-T0008.002) 
  & \cite{shadowhammer1}, \cite{shadowhammer2} \\

\midrule

Delivery (UEFI-TA9003) 
  & Exploit Supply Chain (UEFI-T0009) 
  & Software Supply Chain (UEFI-T0009.001) 
  & \cite{shadowhammer3}, \cite{shadowhammer4}, \cite{cosmicstrand1}, \cite{lojax1} \\

  & 
  & Hardware Supply Chain (UEFI-T0009.002) 
  & \cite{cosmicstrand1}, \cite{moonbounce2} \\

  & Compromise Remote Update (UEFI-T0010) 
  & 
  & \cite{lojax1}, \cite{shadowhammer3} \\

  & Physical Attacks (UEFI-T0011) 
  & 
  & \cite{SonicScrewDriver}, \cite{thunderstrike1}, \cite{thunderstrike2}, \cite{vectoredk1} \\

  & Network Attacks (UEFI-T0012) 
  & 
  & \cite{trickboot3}, \cite{moonbounce2}, \cite{DerStarke2} \\

\midrule

Exploitation (UEFI-TA9004) 
  & Overflow Vulnerability (UEFI-T0013) 
  & 
  & \cite{boothole1}, \cite{blacklotus1} \\

  & UEFI Shell Exploitation (UEFI-T0014) 
  & 
  & \cite{uefi_shell_spec} \\

  & DMA Attacks (UEFI-T0015) 
  & 
  & \cite{IOMMU_DMA}, \cite{SonicScrewDriver} \\

  & Code Execution Attacks (UEFI-T0016) 
  & Malicious Driver Injection (UEFI-T0016.001) 
  & \cite{peibackdoor}, \cite{lojax3}, \cite{moonbounce2} \\

  & 
  & Malicious Boot Loader Injection (UEFI-T0016.002) 
  & \cite{Dreamboot1}, \cite{blacklotus1}, \cite{especter1} \\

  & 
  & Malicious Payload Injection (UEFI-T0016.003) 
  & \cite{lojax3}, \cite{mosaicregressor2}, \cite{vectoredk1} \\

  & Exploit UEFI Services (UEFI-T0017) 
  & 
  & \cite{uefi_specs_2_10} \\

  & UEFI Image Tampering (UEFI-T0018) 
  & Direct Flashing (UEFI-T0018.001) 
  & \cite{moonbounce2}, \cite{cosmicstrand1}, \cite{vectoredk1} \\

  & 
  & Malicious Remote Update (UEFI-T0018.002) 
  & \cite{shadowhammer2}, \cite{lojax1} \\

\midrule

Persistence (UEFI-TA9005) 
  & Platform Key Abuse (UEFI-T0019) 
  & 
  & \cite{blacklotus1} \\

  & Bootkit Installation (UEFI-T0020) 
  & 
  & \cite{vectoredk1}, \cite{lojax1}, \cite{moonbounce2}, \cite{cosmicstrand1}, 
    \cite{especter2}, \cite{blacklotus1} \\

  & UEFI Variable Tampering (UEFI-T0021) 
  & 
  & \cite{vectoredk2}, \cite{mosaicregressor1}, \cite{blacklotus1} \\

  & Hooking UEFI Services (UEFI-T0022) 
  & 
  & \cite{moonbounce2}, \cite{cosmicstrand2}, \cite{mosaicregressor2}, \cite{especter2} \\

\midrule

Defense Evasion (UEFI-TA9006) 
  & Code Obfuscation (UEFI-T0023) 
  & 
  & \cite{finspy2}, \cite{blacklotus1} \\

  & Payload Removal (UEFI-T0024) 
  & 
  & \cite{lojax3} \\

  & Subvert UEFI Security Controls (UEFI-T0025) 
  & Secure Boot Bypass (UEFI-T0025.001) 
  & \cite{blacklotus1}, \cite{finspy1} \\

  & 
  & UEFI Security Policy Alteration (UEFI-T0025.002) 
  & \cite{blacklotus1} \\

  & 
  & Code Signing Certificate Abuse (UEFI-T0025.003) 
  & \cite{shadowhammer2} \\

  & 
  & Bypassing UEFI Password (UEFI-T0025.004) 
  & \cite{eclypsium_spi_write_protections} \\

\midrule

Discovery (UEFI-TA9007) 
  & Hooking UEFI API/Services (UEFI-T0026) 
  & 
  & \cite{moonbounce2}, \cite{cosmicstrand2}, \cite{especter1} \\

  & System Architecture Mapping (UEFI-T0027) 
  & 
  & \cite{trickboot3} \\

  & Peripheral Device Discovery (UEFI-T0028) 
  & 
  & \cite{trickboot3} \\

  & Boot Configuration Analysis (UEFI-T0029) 
  & 
  & \cite{blacklotus1} \\

\midrule

Privilege Escalation (UEFI-TA9008) 
  & Manipulating UEFI Variables (UEFI-T0030) 
  & 
  & \cite{blacklotus1}, \cite{uefi_specs_2_10}, \cite{update-runtime-exploitation} \\

  & UEFI Services Exploitation (UEFI-T0031) 
  & 
  & \cite{update-runtime-exploitation} \\

  & Abusing Secure Boot (UEFI-T0032) 
  & 
  & \cite{blacklotus1}, \cite{defense_secure_boot_customization} \\

  & SMM Exploitation (UEFI-T0033) 
  & 
  & \cite{lighteater}, \cite{thinkpwn}, \cite{SMMrootkit} \\

\midrule

Credential Access (UEFI-TA9009) 
  & UEFI Setting Password (UEFI-T0034) 
  & 
  & \cite{nsa_uefi_defensive_practices} \\

  & Leaked Authentic Digital Certificates (UEFI-T0035) 
  & 
  & \cite{shadowhammer2}, \cite{shadowhammer1} \\

  & Digital Certificate and Key Extraction (UEFI-T0036) 
  & 
  & \cite{shadowhammer1} \\

\midrule

Collection (UEFI-TA9010) 
  & UEFI Variable Enumeration (UEFI-T0037) 
  & 
  & \cite{sarvepalli2023securing}, \cite{uefi_specs_2_10} \\

  & Configuration Table Enumeration (UEFI-T0038) 
  & 
  & \cite{uefi_specs_2_10} \\

  & Physical Memory Dump (UEFI-T0039) 
  & 
  & \cite{lighteater} \\

  & UEFI Image Extraction (UEFI-T0040) 
  & 
  & \cite{uefiSPI} \\

  & Network Configuration and Credentials (UEFI-T0041) 
  & 
  & \cite{uefi_specs_2_10} \\

  & UEFI Logs (UEFI-T0042) 
  & 
  & \cite{bitlocker}, \cite{uefi_specs_2_10}, \cite{sarvepalli2023securing} \\

\end{longtable}
\end{tiny}
\flushbottom

\newpage
\begin{landscape}
\begin{figure*}
    \centering
        \caption{Proposed UEFI MITRE ATT\&CK-like framework}
    \label{fig:UEFI_MITRE_framework}
    \includegraphics[width=1.00\linewidth]{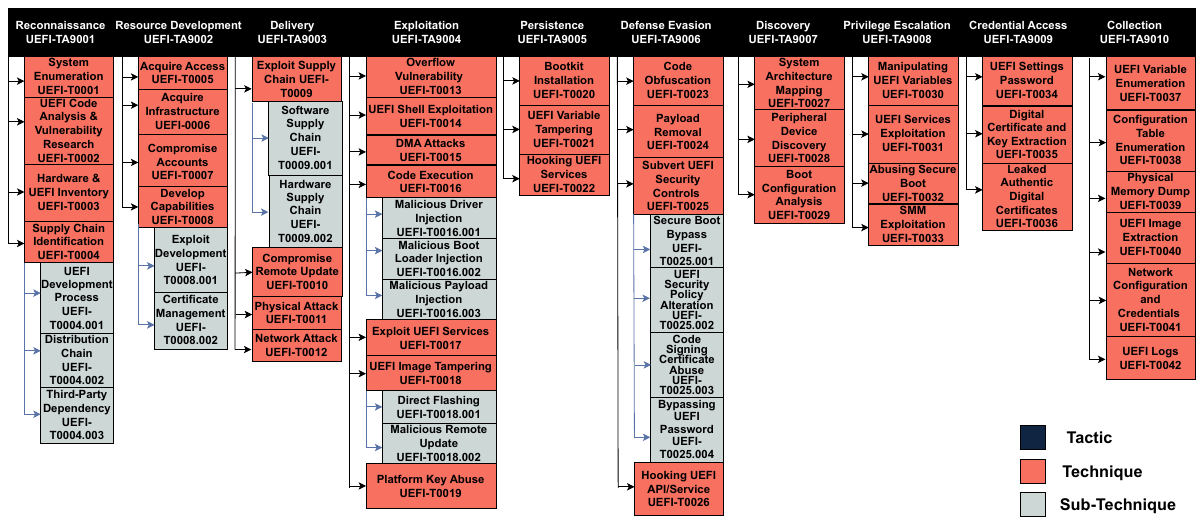}    
\end{figure*}
\end{landscape}

\new{\subsection{Taxonomy Construction Methodology} \label{sec:taxonomy_methodology}}

\new{The taxonomy was derived through a systematic, evidence-driven process grounded in the attack corpus described in Section~\ref{sec:uefi_attacks}. The construction process followed four sequential steps.}

\new{\noindent \textbf{Step 1: Systematic Attack Review.}
All publicly documented UEFI-related attacks and proof-of-concept implementations available at the time of study were reviewed systematically, as summarized in Table~\ref{tab:attacks_table}. 
For each attack, the following dimensions were extracted: the initial access vector, the exploitation mechanism, the persistence technique, the defense evasion approach, and the post-exploitation objectives. 
This extraction was performed directly from primary sources, including academic publications, security vendor reports, and publicly available technical analyses, ensuring that the taxonomy reflects observed real-world behavior rather than theoretical attack models.}

\new{\noindent \textbf{Step 2: Pattern Identification and Grouping.} Extracted attack steps were analyzed to identify recurring patterns across multiple attacks. 
Steps exhibiting similar objectives and mechanisms were grouped into candidate techniques. 
Candidate techniques that shared a common adversarial goal at a higher level of abstraction were further grouped into candidate tactics. 
This bottom-up grouping process was guided by the MITRE ATT\&CK framework~\cite{mitre} as a structural reference, ensuring compatibility with existing security tooling and terminology while remaining grounded in UEFI-specific evidence.}

\new{\noindent \textbf{Step 3: Inclusion and Exclusion Criteria.} 
A tactic or technique was included in the taxonomy if it was supported by the reviewed UEFI attack corpus and represented a behavior that was meaningfully tied to the UEFI or firmware attack surface. Behaviors that may appear in UEFI-related campaigns but are not themselves firmware-specific were not modeled as standalone taxonomy elements unless they played a distinct role in enabling, targeting, or exploiting the UEFI environment. This design choice explains why the proposed framework contains 10 tactics rather than mirroring all 14 tactics in the full MITRE ATT\&CK enterprise framework.}

\new{\noindent \textbf{Step 4: Iterative Review and Refinement.} The resulting taxonomy was subjected to iterative review against the full attack corpus to verify that the documented attacks could be consistently represented using the proposed tactics and techniques. Where initial groupings were found to be ambiguous, technique boundaries were refined to improve consistency and preserve a clear UEFI-specific focus. The taxonomy was further reviewed against the UEFI specification and related security literature to verify that the defined techniques corresponded to distinct, technically meaningful stages of UEFI exploitation. We acknowledge that the taxonomy reflects the state of publicly documented attacks at the time of study and may not capture attack techniques that have not yet been publicly disclosed. We therefore present it as a living framework and invite the UEFI security research community to adopt, validate, and extend it as the threat landscape evolves.}

\subsection{Mapping UEFI Attacks to MITRE-like TTPs (Tactics, Techniques and Procedures) }
\label{subsec:UEFI_taxonomy}
In this section, we discuss the TTPs within the UEFI MITRE ATT\&CK-like framework.
\new{The tactic and technique identifiers used in this framework follow a numbering scheme beginning at 9001 (for tactics) and T0001 (for techniques) to avoid conflicts with existing MITRE ATT\&CK enterprise identifiers, which use four-digit codes in overlapping ranges. This ensures that the UEFI-specific framework can be used alongside existing MITRE frameworks without identifier collision.}
\begin{itemize}
    \item 
\noindent\textbf{Reconnaissance (UEFI-TA9001):}
During reconnaissance, a threat actor gathers information about the target. Within the context of UEFI, this process can include the following:
\begin{itemize}
    \item 

\noindent\textbf{System Enumeration (UEFI-T0001)}: Attackers aim to identify the target system's UEFI firmware version, vendor, and specific model \cite{trickboot3}. This information helps them understand the firmware's features, potential vulnerabilities, and available exploits.
\item 
\noindent \textbf {Firmware Code Analysis and Vulnerability Research (UEFI-T0002):} Threat actors analyze UEFI firmware code to identify potential security weaknesses, such as known issues in outdated firmware versions \cite{shadowhammer1}, known vulnerabilities \cite{blacklotus1}, or misconfigurations \cite{lojax1}. 
\item 
\noindent \textbf{Hardware and Firmware Inventory (UEFI-T0003):} Threat actors gather information about the hardware components integrated into the system, such as the motherboard, chipset, storage devices, and network adapters. They also identify the firmware versions associated with these components, as outdated drivers can be potential entry points for exploitation \cite{trickboot2}.

\item 
\noindent \textbf{Supply Chain Identification (UEFI-T0004):} Threat actors analyze the supply chain for potential security weaknesses at various stages, from development to distribution. In this context, the threat actors do any of the following: 
        \begin{itemize}
            \item \noindent\textbf{\textit{Firmware Development Process (UEFI-T0004.001)}:} Gather information about the firmware development and integration process, tools, methodologies, and security practices employed by the IBV/ODM/OEM.
            \item \noindent\textbf{\textit{Distribution Chain (UEFI-T0004.002)}:} Analyze the security measures deployed by vendors during manufacturing, development, transportation, storage, and installation of the UEFI. 
            \item \noindent\textbf{\textit{Third-Party Dependencies (UEFI-T0004.003)}:} Analyze the dependencies and third-party hardware/software used in the system to identify vulnerabilities in those components that can be exploited to compromise the integrity of the UEFI.
        \end{itemize}
\end{itemize}
\item 
\noindent \textbf{Resource Development (UEFI-TA9002):}
During the resource development stage, threat actors use the information obtained during reconnaissance to craft the tools and means needed to achieve their campaign goals. In this context, they do the following:
\begin{itemize}
    \item 
    \noindent\textbf{Acquire Access (UEFI-T0005):} The adversary purchases access to the target via access brokers~\cite{trickboot1}.
    \item 
    \noindent \textbf{Acquire Infrastructure (UEFI-T0006):} Adversaries establish infrastructure such as C\&C servers to deliver payloads, receive instructions, or maintain persistence within targeted systems~\cite{mosaicregressor2,lojax3}. 
    \item         
    \noindent \textbf{Compromise Accounts (UEFI-T0007):} Attackers compromise accounts with privileged access to jeopardize the security and integrity of UEFI \cite{shadowhammer1}. 
    \item        
    \noindent \textbf{Develop Capabilities (UEFI-T0008):} Attackers develop capabilities, such as exploits, to carry out malicious activities. Those capabilities include the following sub-techniques: 
            \begin{itemize}
                \item \noindent\textbf{\textit{Exploit Development (UEFI-T0008.001)}:} Attackers create or obtain exploit code to target specific UEFI vulnerabilities, such as customizing a UEFI DXE driver~\cite{paloalto_glupteba_malware}.
                \item \noindent\textbf{\textit{Certificate Management (UEFI-T0008.002)}:} Attackers acquire authentic digital certificates from vendors or generate their own to sign malicious UEFI code \cite{shadowhammer1}.  
            \end{itemize}
\end{itemize}
    \item 
    \noindent \textbf{Delivery (UEFI-TA9003)} \new{During this stage, the threat actor attempts to introduce, transmit, or position the exploit or malicious component into the UEFI or pre-OS attack surface. In this taxonomy, the Delivery tactic is limited to UEFI-specific exploit delivery paths, such as supply-chain compromise, remote firmware update paths, physical access, or network-reachable firmware interfaces. More general command-and-control activity is outside the scope of this UEFI-specific framework and is therefore not modeled as a separate tactic here.} This includes: 
        \begin{itemize}
        \item 
        \noindent \textbf{Exploit Supply Chain (UEFI-T0009):} Attackers compromise supply-chain components of UEFI systems to deliver a malicious payload to the target system. This can be accomplished using the following sub-techniques:
        \begin{itemize}
            \item \noindent\textbf{\textit{Software Supply Chain (UEFI-T0009.001)}:} Attackers hijack the UEFI firmware development and distribution mechanism
        ~\cite{cosmicstrand1,shadowhammer3}. 
            \item \noindent\textbf{\textit{Hardware Supply Chain (UEFI-T0009.002)}: } Attackers infiltrate the hardware manufacturing process to compromise the hardware components such as the motherboard (SPI chip) and 3rd party components, ensuring their malicious code (e.g., DXE driver) is executed when the hardware is initiated.
        \end{itemize}
    \item     
    \noindent \textbf{Compromise Remote Update (UEFI-T0010):} Attackers attempt to exploit vulnerabilities in the firmware update mechanism provided by the UEFI. An attacker can trick the system into accepting and applying a malicious update, delivering their payload into the UEFI~\cite{shadowhammer4}. 
      \item 
        \noindent \textbf{Physical Access Attack (UEFI-T0011):} Attackers gain unauthorized physical access to the target system to modify the UEFI firmware by tampering with the SPI flash chip or other storage devices where the firmware is stored. Such an attack can typically be carried out by an insider threat~\cite{SonicScrewDriver,thunderstrike1,vectoredk1}.
        \item 
        \noindent \textbf{Network Access Attacks (UEFI-T0012): } The threat actor initiates communication with a network-connected target to gain access~\cite{DerStarke2,trickboot3,moonbounce2}.
 \end{itemize}
 \item 
   \noindent \textbf{Exploitation (UEFI-TA9004):} During this stage, the threat actor attempts to exploit the target system by running malicious code or using native capabilities (e.g., "living off the land" attacks).
\begin{itemize}
    \item 
       \noindent \textbf{Overflow Vulnerability (UEFI-T0013):} Vulnerabilities can occur when inputs to UEFI system components are not properly validated, or when buffer limits are not enforced, causing data to overflow into adjacent memory regions. Attackers gain unauthorized access to such memory regions and misuse the sensitive information gathered from those regions~\cite{boothole1}. 
       \item 
       \noindent \textbf{UEFI Shell Exploitation (UEFI-T0014):} If the UEFI implementation supports a shell environment, it can be exploited by the attackers to execute arbitrary UEFI shell scripts or commands~\cite{uefi_shell_spec}.
       \item 
       \noindent \textbf{Direct Memory Access (DMA) Attacks (UEFI-T0015):} Attackers exploit the DMA capabilities (the ability to access memory directly) of peripheral devices such as network interface cards, graphics cards, and storage controllers to compromise UEFI firmware or the underlying operating system. This leads to malicious code injection and privilege escalation~\cite{IOMMU_DMA}.
       \item
       \noindent \textbf{Code Execution (UEFI-T0016):} Code injection gives attackers the ability to execute arbitrary code on the system by exploiting vulnerabilities in UEFI firmware or associated components, using the following sub-techniques: 
       \begin{itemize}
            \item \noindent\textbf{\textit{Malicious Driver Injection (UEFI-T0016.001)}:} Attackers can inject and execute a malicious driver during the DXE phase of booting. This allows attackers to gain control over the system very early, potentially compromising UEFI security and integrity ~\cite{thunderstrike2,peibackdoor,lojax3}. 
            \item \noindent\textbf{\textit{Malicious Boot Loader Injection (UEFI-T0016.002)}:} Attackers modify or replace the legitimate bootloader with a malicious one. The bootloader loads the operating system (OS) into memory and initiates system startup. Injecting a malicious bootloader can give attackers control over the boot process ~\cite{Dreamboot2,boothole1,finspy2,blacklotus1}. 
            \item \noindent\textbf{\textit{Malicious Payload Injection (UEFI-T0016.003)}:} Attackers inject and execute malicious code within the UEFI firmware environment through custom code or firmware updates~\cite{Dreamboot1,thunderstrike1,boothole1}.
        \end{itemize}
        \item 
        \noindent \textbf{Exploit UEFI Services (UEFI-T0017):} Threat actors exploit various UEFI services to carry out attacks, such as EFI\_ BOOT\_SERVICES to exploit various boot-related functions (e.g., loading bootloaders and operating systems), EFI\_VARIABLE\_SERVICES to modify UEFI variables (potentially altering system behavior or disabling security features), EFI\_FILE\_SERVICES to manipulate or replace critical UEFI firmware files, EFI\_TEXT\_INPUT\_PROTOCOL to inject malicious commands or bypass security measures, and EFI\_NETWORK\_INTERFACE\_IDENTIFIER\_PROTOCOL to launch remote UEFI attacks, such as firmware updates over the network~\cite{uefi_specs_2_10}.
        \item 
        \noindent \textbf{UEFI Image Tampering (UEFI-T0018):} Tampering with the firmware image is used to establish persistence, bypass UEFI security features, create hidden backdoors within the firmware, provide unauthorized access to the system, or compromise its security. The following sub-techniques are observed at this stage:
        \begin{itemize}
            \item \noindent\textbf{\textit{Physical Access (UEFI-T0018.001)}:} Attackers replace the firmware with a trojanized version by writing the malicious firmware directly to the SPI chip, typically using specialized software or tools provided by the hardware manufacturer~\cite{moonbounce2}.
            \item \noindent\textbf{\textit{Malicious Remote Update (UEFI-T0018.002)}:} Attackers exploit vulnerabilities in the UEFI remote update process to inject malicious firmware updates by manipulating firmware update servers or using social engineering tactics to trick users into applying the update ~\cite{mosaicregressor2,shadowhammer2}.
        \end{itemize}
\item 
        \noindent \textbf{Platform Key Abuse (UEFI-T0019):} Attackers sign malicious UEFI firmware, bootloaders, or drivers with a compromised platform key, making them appear legitimate and trusted by the UEFI. 
\end{itemize}
\item 
   \noindent \textbf{Persistence (UEFI-TA9005)} These techniques ensure that the threat actor maintains control of the infected system, even after a reboot. A threat actor accomplishes persistence in several ways:
\begin{itemize}
    \item 
        \noindent \textbf{Bootkit Installation (UEFI-T0020):} Establish persistence on a compromised system by embedding malicious code within the boot process~\cite{vectoredk1}, ensuring it executes each time the system starts up. 
        \item 
        \noindent \textbf{UEFI Variable Tampering (UEFI-T0021):} UEFI variable tampering ensures that the attacker's malicious configurations or code remain effective across multiple system reboots. Attacker does so by changing the variables containing individual boot entries and defining the paths to bootloader files, the order in which boot devices are accessed during the boot process, PK and KEK variables, etc~\cite{vectoredk2,mosaicregressor1}.
        \item 
        \noindent \textbf{Hooking UEFI Services (UEFI-T0022):} UEFI runtime services provide various functions accessible even after the operating system has been booted. Attackers modify or "hook" these services by injecting malicious code or modifying existing service handlers. Handlers such as GetVariable and SetVariable are targeted to manipulate UEFI variables related to boot order, firmware settings, or security policies \cite{moonbounce2}. LocateProtocol and LocateHandle are used to access specific firmware interfaces or to locate functions or resources. InstallProtocolInterface is exploited to introduce malicious code into protocols or interfaces. AllocatePages and FreePages allocate memory for code or data structures, ensuring persistence, while RaiseTPL is used to gain higher privileges by raising the Task Priority Level~\cite{uefi_specs_2_10,mosaicregressor2,especter2}.
\end{itemize}
\item 
    \noindent \textbf{Defense Evasion (UEFI-TA9006)} 
\begin{itemize}
    \item 
        \noindent \textbf{Code Obfuscation (UEFI-T0023):} Attackers obfuscate malicious UEFI code to make it more difficult for defenders to analyze and identify its purpose. Obfuscation techniques can include encryption, encoding, or packing the code~\cite{finspy2,blacklotus1}.
        \item 
        \noindent \textbf{Payload Removal (UEFI-T0024):} Attackers design their malware to remove malicious files after infecting the target~\cite{lojax3}.
        \item 
        \noindent \textbf{Subvert UEFI Security Controls (UEFI-T0025):} UEFI firmware includes various security mechanisms designed to protect the system against unauthorized access, malware, and other threats~\cite{especter2,blacklotus1}. 
        The attackers exploit vulnerabilities to bypass or undermine the security features such as secure boot, capsule update, and code-signing certificate using the following techniques:
        \begin{itemize}
         \item \noindent \textbf{Secure Boot Bypass (UEFI-T0025.001):} Secure Boot ensures that only trusted and signed firmware, bootloaders, and drivers are executed during the boot process. Vulnerabilities in Secure Boot implementation enables threat actors to load and execute unsigned or malicious code, bypassing the integrity checks performed by Secure Boot~\cite{finspy1,blacklotus1}.
         \item  \noindent\textbf{UEFI Security Policy Alteration (UEFI-T0025.002):} Attackers perform unauthorized modifications or deletions of the UEFI security policies and settings related to Secure Boot, UEFI Secure Flash, platform key management, etc., to escalate privileges, establish persistence, or create a backdoor~\cite{blacklotus1}. 
         \item \noindent\textbf{Code Signing Certificate Abuse (UEFI-T0025.003):} Code signing verifies the authenticity and integrity of software by attaching a digital signature. With a stolen or fraudulent code signing certificate, attackers can sign malicious UEFI firmware, drivers, or other components that are executed during the boot process or firmware update and are considered legitimate due to their valid digital signatures~\cite{shadowhammer2}.
         \item \noindent \textbf{Bypassing UEFI Password (UEFI-T0025.004):} Some systems allow users to set passwords to protect UEFI settings. Attackers use multiple techniques, such as dictionary/brute force attacks, keylogging, or social engineering, to bypass or recover these passwords, gaining administrative access and control over firmware settings~\cite{eclypsium_spi_write_protections}. 
        \end{itemize}
        \item 
        \noindent \textbf{Hooking UEFI API/Service (UEFI-T0026):} Attackers hook UEFI services or redirect UEFI API calls to their own code. This allows them to intercept and manipulate API calls and evade security controls~\cite{especter1,moonbounce2,cosmicstrand2}. 
\end{itemize}
\item \textbf{Discovery (UEFI-TA9007):} 
\begin{itemize}
    \item 
        \noindent \textbf{System Architecture Mapping (UEFI-T0027):} Attackers aim to identify the architecture and components of the target system, including the motherboard, chipset, processors, memory modules, storage devices, and network interfaces. This information helps them understand the system's capabilities, potential attack vectors, and areas of focus for post-exploitation~\cite{trickboot3}.
        \item 
        \noindent \textbf{Peripheral Device Discovery (UEFI-T0028):} Attackers identify peripheral devices connected to the target system, such as USB devices, network adapters, or storage devices. This helps them understand the potential post-exploitation avenues associated with these devices. 
        \item 
        \noindent \textbf{Boot Configuration Analysis (UEFI-T0029):} Attackers explore the boot configuration settings in UEFI firmware, such as the boot order, boot options, and boot manager entries. 
        This involves studying the boot process and associated components to identify weak points or misconfigurations. This information can be useful for post-exploitation manipulation of the boot process or alteration of boot parameters~\cite{blacklotus1}. 
 \end{itemize}
 \item 
    \noindent \textbf{Privilege Escalation (UEFI-TA9008):}
 \begin{itemize}
     \item 
     \noindent \textbf{Manipulating UEFI Variables (UEFI-T0030):} UEFI variables store system settings and configurations, including the security policy. Attackers manipulate these variables to change security settings, boot order, or other critical configurations, thereby escalating their privileges within the firmware environment~\cite{uefi_specs_2_10}. 
        \item 
        \noindent \textbf{UEFI Service Exploitation (UEFI-T0031):} Manipulating UEFI services, such as UEFI variable services, UEFI handles, and task-level priority, allows attackers to gain unauthorized control over the boot process and system configuration~\cite{update-runtime-exploitation}. At this elevated access level, attackers can execute their code without any restrictions.
        \item 
        \noindent \textbf{Abusing Secure Boot (UEFI-T0032):} By exploiting vulnerabilities in secure boot, attackers can inject unsigned code at the firmware level. This results in privilege escalation at the OS level, providing attackers with root-level access~\cite{defense_secure_boot_customization}. 
        \item 
        \noindent \textbf{SMM Exploitation (UEFI-T0033):} Attackers exploit vulnerabilities in SMM code to gain low-level control of the system, effectively escalating their privileges within the firmware~\cite{SMMrootkit}. 
    
\end{itemize}
\item 
    \noindent \textbf{Credential Access (UEFI-TA9009):}
\begin{itemize}
    \item 
        \noindent \textbf{UEFI Settings Password (UEFI-T0034):} UEFI implementations allow users to set passwords to protect UEFI settings. Attackers attempt to extract or bypass these passwords to gain unauthorized access to UEFI settings and credentials~\cite{nsa_uefi_defensive_practices}. 
        
        \item 
        \noindent \textbf{Leaked Authentic Digital Certificates (UEFI-T0035):} Leaked digital certificates, particularly those used for code signing or secure communication, can be exploited by attackers. 
        If attackers gain access to these certificates, they can sign malicious firmware or bootloader components, making them appear legitimate during the Secure Boot verification process~\cite{shadowhammer2}. 
        \item 
        \noindent \textbf{Digital Certificate and Key Extraction (UEFI-T0036):} If digital certificates or encryption keys are stored in UEFI firmware for secure communication or firmware updates, attackers attempt to extract these credentials for later use~\cite{shadowhammer1}. 
        \end{itemize}
        \item 
    \noindent \textbf{Collection (UEFI-TA9010):}
\begin{itemize}
        
        \item 
        \noindent \textbf{UEFI Variable Enumeration (UEFI-T0037):} Attackers enumerate UEFI variables to gather information about system settings, configurations, or stored data~\cite{sarvepalli2023securing}. 
        \item 
        \noindent \textbf{Configuration Table Enumeration (UEFI-T0038):} UEFI firmware uses configuration tables to store information about system hardware and resources. 
        Attackers enumerate these tables to gather details about the system's hardware components and configurations~\cite{uefi_specs_2_10}. 
        \item 
        \noindent \textbf{Physical Memory Dump (UEFI-T0039):} Attackers with access to a system's physical memory perform memory dumps to collect data stored in RAM. This data can include UEFI variables, SMM data, or other sensitive information present in memory during the boot process. 
        \item 
        \noindent \textbf{Firmware Image Extraction (UEFI-T0040):} Attackers attempt to extract the UEFI firmware image from the system for analysis or reverse engineering. 
        \item 
        \noindent \textbf{Network Configuration and Credentials (UEFI-T0041):} UEFI firmware stores network-related configurations and credentials for services like PXE (Preboot Execution Environment) boot or remote management. Attackers extract these credentials for unauthorized network access~\cite{uefi_specs_2_10}. 
        \item         
        \noindent \textbf{UEFI Logs (UEFI-T0042):} Attackers can retrieve measured boot logs (if enabled) or records related to UEFI updates to gather information about previous updates, vulnerabilities, or changes in the firmware~\cite{bitlocker}. 
        \end{itemize}
\end{itemize}

\section{Analysis of Existing Countermeasures}
\label{sec:uefi-countermeasures}
As the threat landscape evolves, the importance of robust countermeasures to safeguard UEFI systems is growing. 
Based on the TTP mapping (see Section~\ref{sec:uefi-mitre}), we identified existing countermeasures to defend UEFI against these threats and, where relevant, we identified gaps and propose additional countermeasures. 
Consistent with the literature, we have broadly classified the countermeasures into categories: prevention, detection, mitigation, and remediation. In line with NIST CSF~\cite{nist}, we have also addressed the "identification" category. Preventive, detective, mitigation, and remediation controls are designed to raise the cost of an attack, while identification controls are designed to understand which organizational assets need protection. However, it is important to remember that the level of abstraction in this paper differs: instead of focusing on the organizational security level (as per NIST CSF), we address the specifics of UEFI. For completeness, and inspired by this category, we mapped the various components of UEFI, as shown in the background section. The countermeasures in the other categories (e.g., prevention, detection, mitigation, and remediation) are summarized in an MITRE-like framework in Figure \ref{fig:UEFI_Countermeasures}. A brief description of the countermeasures follows; for a  detailed description of each countermeasure, refer to the references in Table \ref{tab:uefi-countermeasures}.

\begin{tiny}
\begin{longtable}{>{\raggedright\arraybackslash}p{1.5cm} >{\raggedright\arraybackslash}p{3cm} >{\raggedright\arraybackslash}p{6cm} >{\raggedright\arraybackslash}p{2cm}}
\caption{Countermeasures for UEFI Security}\label{tab:uefi-countermeasures} \\

\toprule
\textbf{Category} & \textbf{Sub-type} & \textbf{List of Countermeasures} & \textbf{Reference} \\
\midrule
\endfirsthead

\multicolumn{4}{c}{\tablename\ \thetable\ -- \textit{Continued from previous page}} \\
\toprule
\textbf{Type} & \textbf{Sub-type} & \textbf{List of Countermeasures} & \textbf{Reference} \\
\midrule
\endhead

\midrule
\multicolumn{4}{r}{\textit{Continued on next page}} \\
\endfoot

\bottomrule
\endlastfoot

Preventive & UEFI Hardening PM-01 
  & Use Secure Boot 
  & \cite{nsa_uefi_defensive_practices}, \cite{defense_secure_boot_customization}, \cite{uefi_secure_boot_insyde} \\
 & & Implement UEFI image signing 
   & \cite{nsa_uefi_defensive_practices}, \cite{defense_secure_boot_customization} \\
 & & BIOS integrity check 
   & \cite{nsa_uefi_defensive_practices}, \cite{nist_sp800_147} \\
 & & Set flash descriptor region to Read-Only 
   & \cite{flashdescriptor}, \cite{eclypsium_spi_write_protections} \\
 & & Disable unnecessary SMM features 
   & \cite{flashdescriptor}, \cite{SMMrootkit} \\
 & & Encrypt sensitive memory areas 
   & \cite{uefi_memory_protection} \\
 & & Memory access control 
   & \cite{IOMMU_DMA}, \cite{uefi_memory_protection} \\
 & & Boot order verification 
   & \cite{nsa_uefi_defensive_practices}, \cite{best-practice-sig-sec-devp} \\
 & & Memory integrity verification 
   & \cite{nsa_uefi_defensive_practices}, \cite{uefi_memory_protection} \\
 & & Regularly update DBX database 
   & \cite{eclypsiumsecureboot}, \cite{microsoft_boot_manager_revocations} \\
 & & Lock down SPI flash memory access 
   & \cite{flashdescriptor}, \cite{eclypsium_spi_write_protections} \\
 & & Full-disk encryption 
   & \cite{bitlocker} \\
 & & Disable legacy protocols and services 
   & \cite{nsa_uefi_defensive_practices} \\
\midrule

Preventive & Supply Chain Security PM-02 
  & Assess and audit vendors in the UEFI supply chain 
  & \cite{blackhat_uefi_safeguarding}, \cite{uefi_virtual_summit_2022} \\
 & & Use secure, traceable supply chain method 
   & \cite{SBOM-supplychain}, \cite{uefi_virtual_summit_2022} \\
 & & Use components from trusted sources only 
   & \cite{blackhat_uefi_safeguarding}, \cite{uefi_secure_boot_insyde} \\
 & & 3rd party component integrity check 
   & \cite{uefi_vulnerability_management}, \cite{blackhat_uefi_safeguarding} \\
 & & Incorporate SBOM into SSDLC 
   & \cite{uefi_virtual_summit_2022}, \cite{SBOM-supplychain}, \cite{uefi_sbom_support} \\
\midrule

Preventive & Authentication, Authorization, Access Control PM-03 
  & Key management policy enforcement 
  & \cite{uefi_kms_implementation}, \cite{uefi_key_management_service} \\
 & & Implement robust Authentication Protocols 
   & \cite{UEFI-MFA}, \cite{uefi_plugfest_2014} \\
 & & Access control to UEFI settings and features 
   & \cite{nsa_uefi_defensive_practices}, \cite{uefi_plugfest_2014} \\
 & & Disable USB Boot 
   & \cite{nsa_uefi_defensive_practices} \\
 & & Ensure Physical Security of Ports 
   & \cite{nsa_uefi_defensive_practices}, \cite{eclypsium_evil_maid_attacks} \\
\midrule

Preventive & Update Patch Management PM-04 
  & Disable UEFI Version Rollback 
  & \cite{uefi_plugfest_2012} \\
 & & Patch and Update the UEFI Regularly 
   & \cite{nsa_uefi_defensive_practices}, \cite{uefi_vulnerability_management} \\
 & & Implement UEFI Version Control 
   & \cite{uefi_vulnerability_management}, \cite{spiceworks_version_control} \\
 & & Implement vulnerability management program 
   & \cite{uefi_vulnerability_management}, \cite{rapid7_vulnerability_management} \\
 & & Verify integrity of update tools 
   & \cite{nsa_uefi_defensive_practices}, \cite{uefi_firmware_security_concerns} \\
 & & Check signature before applying UEFI update 
   & \cite{nsa_uefi_defensive_practices}, \cite{defense_secure_boot_customization} \\
 & & MFA for update process 
   & \cite{UEFI-MFA}, \cite{entitle_mfa} \\
\midrule

Preventive & Execution Control PM-05 
  & Software execution policy 
  & \cite{uefi_kms_implementation}, \cite{uefi_specs_2_10} \\
 & & Run only signed drivers 
   & \cite{defense_signed_software_execution}, \cite{uefi_secure_boot_driver_signing} \\
 & & Firmware integrity 
   & \cite{uefi_intel_defense}, \cite{uefi_key_management_service}, 
     \cite{nsa_uefi_defensive_practices}, \cite{uefi_secure_coding_firmware} \\
\midrule

Detective & System Behavior Monitoring DM-01 
  & Monitor SMM access and activities 
  & \cite{IOMMU_DMA}, \cite{SMMrootkit} \\
 & & Monitor system during loading of a driver 
   & \cite{nsa_uefi_defensive_practices}, \cite{flashdescriptor} \\
 & & Scan for unauthorized memory access patterns 
   & \cite{IOMMU_DMA}, \cite{uefi_memory_protection} \\
 & & Monitor for memory patching attempts 
   & \cite{IOMMU_DMA}, \cite{flashdescriptor} \\
 & & Monitor DMA access 
   & \cite{IOMMU_DMA}, \cite{eclypsium_dma_attacks} \\
 & & Monitor unexpected Write operations to the flash descriptor 
   & \cite{flashdescriptor}, \cite{eclypsium_spi_write_protections} \\
 & & Monitor UEFI logs 
   & \cite{nsa_uefi_defensive_practices}, \cite{sarvepalli2023securing} \\
\midrule

Detective & Anomaly Detection DM-02 
  & Monitor for unusual activities 
  & \cite{labunets_efixplorer}, \cite{RelWorkyang2020uefifuzzing} \\
 & & Conduct UEFI static analysis 
   & \cite{binarly_symbolic_execution}, \cite{labunets_efixplorer}, 
     \cite{uefi_firmware_security_concerns} \\
\midrule

Mitigation & Incident Response MM-01 
  & Investigation of Suspicious Activities 
  & \cite{uefi_firmware_security_concerns}, \cite{cisa_cybersecurity_response_playbooks} \\
 & & Analyze security feature bypass 
   & \cite{RelWorkkallenberg2014defeatingsecboot}, \cite{uefi_firmware_security_concerns} \\
 & & Analyze supply chain impact 
   & \cite{blackhat_uefi_safeguarding}, \cite{uefi_virtual_summit_2022} \\
\midrule

Mitigation & Isolation MM-02 
  & Isolate the affected system 
  & \cite{cisa_cybersecurity_response_playbooks}, \cite{uefi_firmware_security_concerns} \\
 & & Disable compromised update paths 
   & \cite{cisa_cybersecurity_response_playbooks}, \cite{nsa_uefi_defensive_practices} \\
 & & Isolate and disable compromise SMM features 
   & \cite{SMMrootkit}, \cite{RelWorkkallenberg2014defeatingsecboot} \\
\midrule

Mitigation & Credential Management MM-03 
  & Revoke compromised credentials 
  & \cite{cisa_uefi_cybersecurity_call_to_action}, \cite{eclypsium_sunburst_firmware_attack}, 
    \cite{microsoft_boot_manager_revocations} \\
 & & Revoke compromised cryptographic keys 
   & \cite{uefi_key_management_service}, \cite{uefi_secure_boot_modern_security}, 
     \cite{microsoft_boot_manager_revocations} \\
\midrule

Remediation & UEFI Restoration RM-01 
  & Reflash UEFI 
  & \cite{uefi_vulnerability_management}, \cite{cisa_cybersecurity_response_playbooks} \\
 & & Reissue cryptographic keys 
   & \cite{uefi_key_management_service}, \cite{uefi_secure_boot_modern_security} \\
 & & Restore credentials 
   & \cite{cisa_cybersecurity_response_playbooks}, \cite{nsa_uefi_defensive_practices} \\

\end{longtable}
\end{tiny}
\flushbottom

\subsection{Preventive Measures}
Preventive countermeasures aim to prevent threat actors from posing a risk to UEFI.  

\noindent
\textbf{UEFI Hardening (PM-01):} Implementing this countermeasure involves several steps, beginning with secure boot, which ensures the integrity of firmware components by preventing the execution of unsigned code (e.g., UEFI drivers and applications) during boot~\cite{nsa_uefi_defensive_practices}. 
UEFI image signing enhances security by authenticating firmware components and safeguarding against unauthorized modifications~\cite{nsa_uefi_defensive_practices}. Cryptographic checks of UEFI updates ensure that only verified updates are applied, maintaining firmware security~\cite{defense_secure_boot_customization}.
Additionally, setting the flash descriptor region to read-only~\cite{flashdescriptor} and disabling unnecessary SMM features~\cite{SMMrootkit} reduce the risk of unauthorized modifications. Encrypting sensitive memory areas further fortifies the system against unauthorized access to data and settings~\cite{uefi_secure_boot_insyde}.
Checks like boot order verification are useful in preventing a variety of attacks~\cite{best-practice-sig-sec-devp}.
Memory integrity verification and memory access control implies implementing mechanisms which prevent unauthorised access to UEFI related structures during boot and run-time.
Full-disk encryption (FDE) is a key method for protecting data by encrypting the entire disk, preventing unauthorized access to the ESP partition, which stores UEFI settings~\cite{bitlocker}.

\noindent
\textbf{Supply Chain Security (PM-02):} Ensuring the security of the UEFI supply chain is crucial. Protecting the UEFI supply chain involves thorough assessment and auditing of vendors to ensure compliance with security standards and UEFI specifications, using secure and traceable software and hardware development and delivery methods~\cite{SBOM-supplychain}, sourcing software and hardware from trusted sources, performing integrity checks on third-party software and hardware components, and incorporating a software bill of materials (SBOM) into the secure software development life cycle (SSDLC) to enhance transparency~\cite{uefi_virtual_summit_2022,blackhat_uefi_safeguarding,uefi_secure_boot_insyde}.

\noindent
\textbf{Access, Authentication \& Authorization Control (PM-03):} This includes restricting access to UEFI settings with password protection to prevent unauthorized configuration modifications. Disabling the ability to boot from USB devices helps protect against boot-level malware. Ensuring the physical security of ports prevents direct access attacks, such as booting from an external device or using hardware keyloggers~\cite{uefi_plugfest_2014}. Enforcing secure key management policies and implementing robust authentication protocols, are crucial for maintaining the security of UEFI systems~\cite{UEFI-MFA,uefi_kms_implementation}.

\noindent
\textbf{Update \& Patch Management (PM-04):} Effective management includes disabling UEFI version rollback to prevent the use of older, vulnerable firmware versions~\cite{uefi_plugfest_2012}, regularly deploying patches to address security issues~\cite{nsa_uefi_defensive_practices}, managing firmware versions by monitoring fleet-wide UEFI versioning and tracking changes with regular assessments, requiring MFA for updates, and implementing a comprehensive vulnerability management program for tracking and remediating vulnerabilities~\cite{uefi_firmware_security_concerns,uefi_vulnerability_management}.

\noindent
    \textbf{Execution Control (PM-05):} This countermeasure involves several controls that restrict code execution in a UEFI setting. For instance, code execution policy enforces rules that allow only pre-approved UEFI-related code (e.g., drivers and applications) to run~\cite{uefi_kms_implementation}. Signed driver enforcement ensures that only authenticated drivers are executed in the UEFI environment~\cite{defense_signed_software_execution}.  Cryptographic methods like digital signatures, hash functions, and certificates should also be used to ensure UEFI firmware integrity~\cite{uefi_specs_2_10,uefi_intel_defense,uefi_key_management_service}.

\noindent

\subsection{Detective Measures}
Detective measures focus on monitoring system behavior to identify potential security breaches or unauthorized activities that may indicate an active attack.

\noindent
\textbf{System Behavior Monitoring (DM-01):} Monitoring activities include checking SMM access, conducting regular UEFI log audits, performing BIOS integrity checks, scanning for unauthorized UEFI/SMM memory access patterns, monitoring UEFI/SMM memory patching attempts, detecting direct memory access (DMA) operations~\cite{IOMMU_DMA}, and observing unexpected write operations to the flash descriptor~\cite{flashdescriptor}.

\noindent 
\textbf{Anomaly Detection (DM-02):} This involves two key approaches: UEFI dynamic analysis to monitor for unusual activities~\cite{labunets_efixplorer}, and UEFI static analysis~\cite{RelWorkyang2020uefifuzzing,binarly_symbolic_execution}. Dynamic analysis scrutinizes UEFI behavior during execution, monitoring run-time activities to detect suspicious or malicious actions. 
In contrast, static analysis examines UEFI firmware without execution, analyzing source code, binary code, and related files to identify potential vulnerabilities and insecure practices. 
Both methods use specialized tools and techniques for UEFI security analysis~\cite{uefi_firmware_security_concerns}.

\subsection{Mitigation Measures}
Mitigation measures aim to address active threats and prevent further exploitation, playing a crucial role in reducing the impact of attacks. 

\noindent
\textbf{Incident Response (MM-01):} This involves investigating suspicious activities within the UEFI environment identified through preventive and detective controls ~\cite{uefi_firmware_security_concerns}. This includes collecting and analyzing system logs, memory dumps, and other digital artifacts to identify the root cause of the incident, assess the extent of the compromise, and gather evidence for further analysis. This step also involves assessing the impact on the supply chain ~\cite{blackhat_uefi_safeguarding}.

\noindent\textbf{Isolation (MM-02):} A key aspect of this countermeasure is isolating affected systems to prevent further infection. Additionally, isolating and disabling affected SMM features is important, as compromised SMM can lead to significant security risks~\cite{SMMrootkit}. Identify and disable compromised update paths to prevent unauthorized or malicious firmware updates. Isolation also involves assessing the integrity of UEFI update mechanisms, such as firmware update utilities, bootloaders, or firmware repositories, and disabling compromised update paths~\cite{cisa_cybersecurity_response_playbooks}. Similarly, analyzing security feature bypasses involves examining the effectiveness of UEFI security mechanisms and identifying potential ways attackers could bypass or exploit these features ~\cite{RelWorkkallenberg2014defeatingsecboot,uefi_firmware_security_concerns}.

\noindent
\textbf{Credential Management (MM-03):} Effective credential management includes revoking compromised credentials and cryptographic keys used in the UEFI environment ~\cite{uefi_key_management_service, eclypsium_sunburst_firmware_attack,uefi_secure_boot_modern_security}. This involves disabling or revoking user accounts, access tokens, or digital certificates that have been compromised or suspected of compromise~\cite{microsoft_boot_manager_revocations,cisa_uefi_cybersecurity_call_to_action}.

\subsection{Remediation Measures}
Effective remediation measures restore impacted systems and the integrity and security of the impacted UEFI environment.

\noindent
\textbf{UEFI Restoration (RM-01):} UEFI restoration involves reflashing firmware to apply the latest updates and patches, addressing known vulnerabilities~\cite{uefi_vulnerability_management,cisa_cybersecurity_response_playbooks, uefi_key_management_service}. 
Re-issuing cryptographic keys for security mechanisms, such as secure boot or encryption, is recommended to invalidate compromised keys. 
Additionally, restoring credentials, including resetting passwords and implementing strong authentication mechanisms, is crucial to safeguarding UEFI system access.

\newpage
\begin{landscape}
\begin{figure*}
\caption{UEFI Countermeasures}
    \centering
    \includegraphics[width=1.0\linewidth]{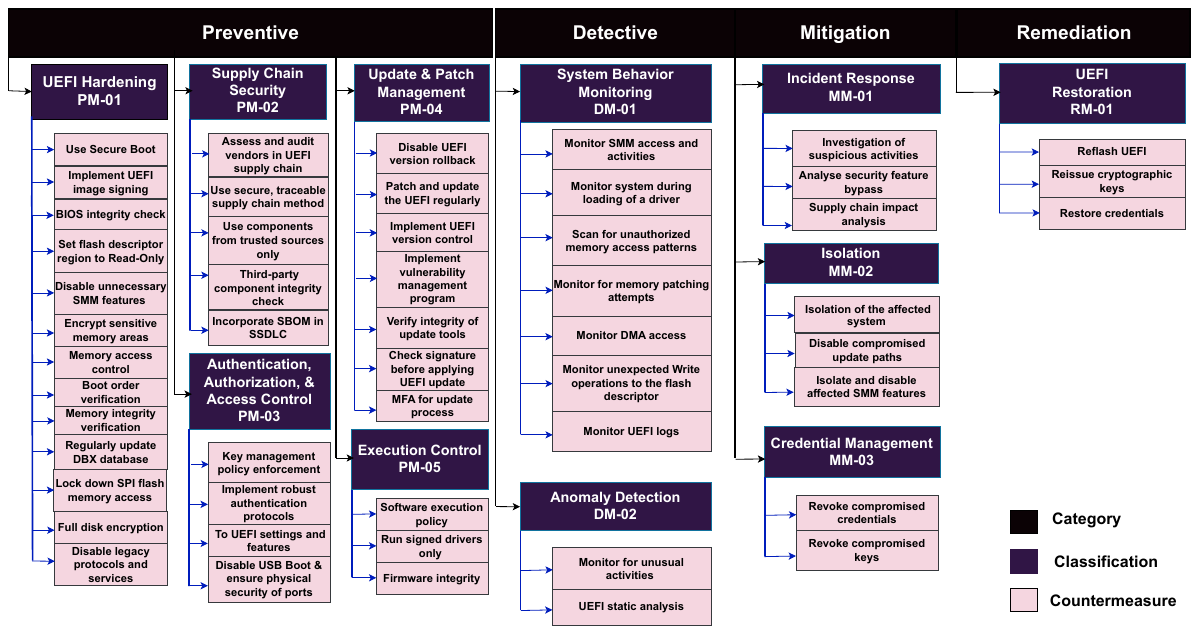}    
     
    \label{fig:UEFI_Countermeasures}
    
\end{figure*}
\end{landscape}

\section{Risk Analysis}
\label{sec:riskanalysis}
We conducted a qualitative risk assessment considering multiple factors 
that influence the overall risk level~\cite{qual-risk}. Integrating 
impact, likelihood, and countermeasure effectiveness allows for a more 
comprehensive understanding of each technique's potential consequences.

\new{\noindent \textbf{Methodology:}
We derived the risk level for each UEFI attack technique as a function of 
impact, likelihood, and effectiveness~\cite{risk-standard800-30} based 
on the following matrices:}

\new{\noindent \textbf{Impact Analysis (I) [Very High, High, Moderate, Low, Very Low]:}}

\noindent The impact of a threat is directly proportional to factors 
such as the nature of the threat, the value and sensitivity of the 
targeted assets, and the extent of the threat's reach within the UEFI 
system. For example, in the context of UEFI, the impact of malicious 
payload injection into a DXE driver is far more severe than UEFI code 
analysis, as payload injection enables immediate and potentially 
far-reaching exploitation. In contrast, UEFI code analysis is a 
precursor to potential threats. Its impact is conditional on the 
subsequent use of discovered information for malicious purposes. We 
conducted an impact analysis for each technique based on the following 
criteria:

\new{
\tiny{
\begin{table}[H]
    \caption{Likelihood Assessment Matrix}
    \centering
    \begin{tabular}{c c p{6.2cm}} 
        \toprule
        Score & Level & Description \\ 
        \midrule
        \priority{100} & Very High & Full resources to execute the attack are at the attacker's disposal \\ 
        \priority{75} & High & The attack can succeed using the acquired resources with little effort \\ 
        \priority{50} & Moderate & Insufficient resources are available to complete the attack \\ 
        \priority{25} & Low & Great effort is required to acquire the resources required to perform this technique \\ 
        \priority{0} & Very Low & Highly unlikely to execute the technique given the limited time and resources \\ 
        \bottomrule
    \end{tabular}
    \label{tab:resource_impact}
\end{table}
}
}

\new{\noindent \textbf{Likelihood Assessment (L) [Very High, High, Moderate, Low, Very Low]:}}

\noindent The likelihood of a risk materializing depends on the UEFI 
attack surface, the threat actor's capability and intent, the presence 
of known vulnerabilities within the UEFI, and the effectiveness of UEFI 
countermeasures. We conducted a likelihood assessment for each of the 
possible techniques:

\new{
\tiny{
\begin{table}[H]
    \caption{Likelihood Assessment Matrix}
    \centering
    \begin{tabular}{c c p{6.2cm}} 
        \toprule
        Score & Level & Description \\ 
        \midrule
        \priority{100} & Very High & Full resources to execute the attack are at the attacker's disposal \\ 
        \priority{75} & High & The attack can succeed using the acquired resources with little effort \\ 
        \priority{50} & Moderate & Insufficient resources are available to complete the attack \\ 
        \priority{25} & Low & Great effort is required to acquire the resources required to perform this technique \\ 
        \priority{0} & Very Low & Highly unlikely to execute the technique given the limited time and resources \\ 
        \bottomrule
    \end{tabular}
    \label{tab:resource_impact}
\end{table}
}
}

\new{\noindent \textbf{Effectiveness Score (E) [Not Applicable, Low, Moderate, High]:}}

\noindent To identify UEFI threats, we examined the attacks and 
vulnerabilities associated with UEFI (see 
Section~\ref{sec:uefi_attacks}). The study resulted in a common pattern 
observed in attacks and we mapped that to MITRE ATT\&CK-like framework 
(refer to Figure~\ref{fig:UEFI_MITRE_framework}). Next, we presented 
countermeasures for each UEFI attack technique (see 
Figure~\ref{fig:UEFI_Countermeasures}). For each countermeasure and 
attack technique pair, we assigned an applicability score (A 0--3) as:

\begin{tiny}

\begin{table}[H]
\caption{Applicability Score Matrix}
\centering
\begin{tabular}{cp{7cm}}
\toprule
\textbf{Score} & \textbf{Description} \\
\midrule
0 & The countermeasure is not applicable \\
\midrule
1 & The countermeasure is useful but has an indirect impact \\
\midrule
2 & The countermeasure is important and quite effective but may not 
cover all aspects to defend the given UEFI attack technique \\
\midrule
3 & The countermeasure is highly effective and plays a direct and 
critical role in ensuring UEFI security \\
\bottomrule
\end{tabular}
\label{tab:applicability_score}
\end{table}
\end{tiny}

\new{\noindent Next, for each UEFI attack technique $t$, we calculate the 
\textit{effectiveness score} $E(t)$ by aggregating the applicability 
scores assigned to the mapped countermeasures. Let $a(t,c)$ denote the 
applicability score assigned to countermeasure $c$ for attack technique 
$t$, according to Table~\ref{tab:applicability_score}, where 
$a(t,c)\in\{0,1,2,3\}$ and $a(t,c)=0$ indicates that the countermeasure 
is not applicable. Let $C_t$ denote the set of applicable mapped 
countermeasures for technique $t$, and let $N_t = |C_t|$ denote the 
number of such mapped countermeasures. The effectiveness score is 
computed as:}

\new{
\begin{equation}
E(t)=\frac{1}{N_t}\sum_{c \in C_t} a(t,c)
\label{eq:effectiveness_score}
\end{equation}
}

\new{\noindent If no applicable mapped countermeasure exists for a given 
technique, we set $E(t)=0$ by definition. This formulation ensures that 
the effectiveness score reflects the average defensive relevance of the 
countermeasures that are actually applicable to the given technique. 
Although the applicability scores themselves are discrete, the 
aggregated effectiveness score $E(t)$ may take non-integer values 
because it is computed as an average. The applicability values are 
expert-assigned ordinal scores intended to approximate the relative 
defensive relevance of a countermeasure to a given UEFI attack 
technique, rather than empirically measured effect sizes. Accordingly, 
the resulting effectiveness values should be interpreted as structured 
comparative indicators for prioritization rather than as exact 
quantitative measurements.}

\new{\noindent We then map $E(t)$ to the qualitative matrix scale as:}

\new{
\begin{table}[H]
\caption{Priority Levels Based on Effectiveness Score}
\centering
\begin{tabular}{cc}
\toprule
\textbf{Effectiveness Tier} & \textbf{E Score Range} \\
\midrule
Low Coverage & 0 -- 1.49 \\
\midrule
Moderate Coverage & 1.5 -- 2.5 \\
\midrule
Strong Coverage & Above 2.5 \\
\bottomrule
\end{tabular}
\label{tab:priority_levels}
\end{table}
}

\new{\noindent The thresholds of 1.5 and 2.5 are grounded in the 
structure of the applicability scale. Since the scale ranges from 0 to 
3, the value 1.5 marks the midpoint between low and moderate coverage, 
while 2.5 marks the upper region in which countermeasures on average 
approach strong, direct coverage. These thresholds therefore divide the 
0--3 range into three semantically meaningful regions corresponding to 
low, moderate, and strong countermeasure coverage.}

\new{\noindent \textbf{Results of the Analysis:} The final qualitative 
risk level is derived from the joint consideration of impact $(I)$, 
likelihood $(L)$, and effectiveness $E(t)$. First, impact and likelihood 
are combined to determine the baseline severity of a technique. In the 
revised analysis, we operationalize this baseline severity using the 
product $I \times L$: techniques with $I \times L \geq 16$ are 
categorized as \textit{High}, techniques with $8 \leq I \times L < 16$ 
are categorized as \textit{Moderate}, and techniques with $I \times L < 
8$ are categorized as \textit{Low}. The effectiveness score then acts as 
an adjustment factor: techniques with weaker countermeasure coverage 
remain in higher risk tiers, whereas techniques with stronger 
countermeasure coverage may be placed in lower tiers despite having 
similar impact and likelihood.}

\new{\noindent For interpretability, we group effectiveness into three 
broad coverage tiers using Table~\ref{tab:priority_levels}. We then 
derive the final risk level according to the rubric in 
Table~\ref{tab:risk_rubric}.}

\new{
\begin{table}[H]
\caption{Rubric for Deriving the Final Risk Level}
\centering
\begin{tabular}{p{4cm}p{4cm}p{2cm}}
\toprule
\textbf{Baseline Severity from $I$ and $L$} & \textbf{Effectiveness Tier} & \textbf{Final Risk} \\
\midrule
High & Low or Moderate Coverage & High \\
High & Strong Coverage & Mid \\
Moderate & Low or Moderate Coverage & Mid \\
Moderate & Strong Coverage & Fair \\
Low & Low or Moderate Coverage & Fair \\
Low & Strong Coverage & Low \\
\bottomrule
\end{tabular}
\label{tab:risk_rubric}
\end{table}
}

\new{\noindent The thresholds used for effectiveness and final risk 
grouping are intended to group techniques into broad comparative risk 
tiers for interpretability. They should therefore be understood as 
presentation-oriented cutoffs rather than exact quantitative 
boundaries.}

\new{\noindent The risk analysis reported in this section is computed over the finalized set of techniques and sub-techniques included in the taxonomy presented in Section~\ref{sec:uefi-mitre}. The results of this analysis, presented in Table~\ref{tab:risk_analysis_results}, indicate that several techniques remain highly exposed despite the presence of multiple relevant countermeasures. In particular, techniques such as malicious boot loader injection, malicious payload injection, malicious driver injection, physical attacks, and SMM exploitation remain in the highest qualitative risk tier. At the same time, a broader set of techniques fall into the mid-risk category, indicating that existing countermeasures provide partial but still incomplete coverage across much of the UEFI attack surface. This finding emphasizes the need for stronger layered defenses, more robust firmware-specific monitoring, and improved protection of the most security-critical stages of the UEFI lifecycle. While many techniques are meaningfully countered, no single control provides complete protection across the full UEFI threat landscape, reinforcing the need for a layered and adaptive security strategy.}

\begin{longtable}{
  c
  >{\RaggedRight\arraybackslash}p{8.2cm}
  c c c c
}
\caption{UEFI Risk Analysis Results}
\label{tab:risk_analysis_results}\\
\toprule
\rowcolor{headerbg}
\textbf{\#} &
\textbf{UEFI Technique} &
\textbf{(I)} & \textbf{(L)} & \textbf{(E)} &
\textbf{Risk} \\
\midrule
\endfirsthead

\toprule
\rowcolor{headerbg}
\textbf{\#} &
\textbf{UEFI Technique} &
\textbf{(I)} & \textbf{(L)} & \textbf{(E)} &
\textbf{Risk} \\
\midrule
\endhead

\midrule
\multicolumn{6}{r}{\small\textit{Continued on next page}} \\
\endfoot

\bottomrule
\endlastfoot

1  & Code Execution – Malicious Boot Loader Injection          & 5 & 4 & 1.92 & \highbadge \\
\rowcolor{rowalt}
2  & Code Execution – Malicious Payload Injection              & 5 & 4 & 1.92 & \highbadge \\
3  & Physical Attacks                                          & 5 & 4 & 2.00 & \highbadge \\
\rowcolor{rowalt}
4  & Code Execution – Malicious Driver Injection               & 5 & 4 & 2.08 & \highbadge \\
5  & SMM Exploitation                                          & 5 & 4 & 2.17 & \highbadge \\
\rowcolor{rowalt}
6  & Subvert UEFI Security Controls – UEFI Security Policy Alteration & 4 & 4 & 1.62 & \highbadge \\
7  & Bootkit Installation                                      & 4 & 4 & 1.79 & \highbadge \\
\rowcolor{rowalt}
8  & Subvert UEFI Security Controls – Secure Boot Bypass       & 4 & 4 & 1.88 & \highbadge \\
\midrule
9  & Exploit UEFI Services                                     & 5 & 3 & 1.50 & \midbadge \\
\rowcolor{rowalt}
10 & Compromise Accounts                                       & 5 & 3 & 2.18 & \midbadge \\
11 & UEFI Image Tampering – Malicious Remote Update            & 5 & 3 & 2.26 & \midbadge \\
\rowcolor{rowalt}
12 & UEFI Image Tampering – Direct Flashing                    & 5 & 3 & 2.43 & \midbadge \\
13 & Hardware and UEFI Inventory                               & 3 & 4 & 1.75 & \midbadge \\
\rowcolor{rowalt}
14 & Network Attacks                                           & 2 & 5 & 1.60 & \midbadge \\
15 & Supply Chain Identification – Distribution Chain          & 3 & 3 & 2.20 & \midbadge \\
\rowcolor{rowalt}
16 & Compromise Remote Update                                  & 5 & 2 & 2.28 & \midbadge \\
17 & UEFI Image Extraction                                     & 3 & 4 & 1.88 & \midbadge \\
\rowcolor{rowalt}
18 & System Enumeration                                        & 3 & 4 & 1.62 & \midbadge \\
19 & Social Engineering                                        & 3 & 4 & 1.75 & \midbadge \\
\rowcolor{rowalt}
25 & UEFI Services Exploitation                                & 4 & 3 & 1.50 & \midbadge \\
27 & Develop Capabilities – Exploit Development                & 3 & 3 & 1.22 & \midbadge \\
\rowcolor{rowalt}
29 & Hooking UEFI API/Services                                 & 3 & 3 & 1.50 & \midbadge \\
30 & Hooking UEFI Services                                     & 3 & 3 & 1.50 & \midbadge \\
\rowcolor{rowalt}
31 & System Architecture Mapping                               & 3 & 3 & 1.50 & \midbadge \\
32 & UEFI Services Exploitation                                & 4 & 3 & 1.50 & \midbadge \\
\rowcolor{rowalt}
33 & Exploit Supply Chain – Hardware Supply Chain              & 3 & 4 & 1.57 & \midbadge \\
34 & Leaked Authentic Digital Certificates                     & 4 & 2 & 1.54 & \midbadge \\
\rowcolor{rowalt}
35 & Network Configuration and Credentials                     & 2 & 5 & 1.56 & \midbadge \\
36 & UEFI Code Analysis and Vulnerability Research             & 3 & 4 & 1.60 & \midbadge \\
\rowcolor{rowalt}
42 & Supply Chain Identification – Distribution Chain          & 3 & 3 & 2.20 & \midbadge \\
43 & Supply Chain Identification – Third-Party Dependency      & 3 & 3 & 1.75 & \midbadge \\
\rowcolor{rowalt}
44 & Overflow Vulnerability                                    & 4 & 3 & 1.75 & \midbadge \\
46 & DMA Attacks                                               & 4 & 3 & 2.00 & \midbadge \\
\midrule
\rowcolor{rowalt}
20 & Boot Configuration Analysis                               & 3 & 3 & 1.17 & \modbadge \\
21 & Hooking UEFI Services                                     & 3 & 3 & 1.50 & \modbadge \\
\rowcolor{rowalt}
22 & System Architecture Mapping                               & 3 & 3 & 1.50 & \modbadge \\
23 & Peripheral Device Discovery                               & 3 & 4 & 1.29 & \modbadge \\
\rowcolor{rowalt}
26 & Abusing Secure Boot                                       & 3 & 3 & 1.27 & \modbadge \\
37 & Code Signing Certificate Abuse                            & 4 & 2 & 1.67 & \modbadge \\
\rowcolor{rowalt}
38 & UEFI Code Analysis and Vulnerability Research             & 3 & 4 & 1.60 & \modbadge \\
39 & Third-Party Dependency                                    & 3 & 3 & 1.75 & \modbadge \\
\rowcolor{rowalt}
40 & Acquire Access                                            & 4 & 2 & 1.82 & \modbadge \\
45 & UEFI Shell Exploitation                                   & 3 & 3 & 1.57 & \modbadge \\
\midrule
\rowcolor{rowalt}
24 & Manipulating UEFI Variables                               & 3 & 2 & 1.40 & \fairbadge \\
41 & Supply Chain Identification – UEFI Development Process    & 3 & 2 & 1.75 & \fairbadge \\
\rowcolor{rowalt}
47 & Platform Key Abuse                                        & 3 & 1 & 1.60 & \fairbadge \\
48 & UEFI Variable Tampering                                   & 3 & 2 & 1.40 & \fairbadge \\
\rowcolor{rowalt}
49 & Code Obfuscation                                          & 2 & 4 & 1.25 & \fairbadge \\
50 & Payload Removal                                           & 2 & 4 & 1.17 & \fairbadge \\
\rowcolor{rowalt}
51 & Digital Certificate and Key Extraction                    & 3 & 2 & 1.40 & \fairbadge \\
52 & UEFI Variable Enumeration                                 & 2 & 2 & 1.00 & \fairbadge \\

\end{longtable}
\new{\section{Limitations}\label{sec:limitations}}

\new{This work makes several contributions to the understanding of UEFI security; however, several limitations should be acknowledged to contextualize the findings appropriately.

\noindent \textbf{Subjectivity of Risk Scores.}
The impact (I) and likelihood (L) scores assigned to each attack 
technique were derived through structured expert judgment informed 
by the reviewed attack corpus, rather than through a formal 
elicitation process such as a Delphi study. While this approach is 
consistent with established applied risk assessment frameworks 
including NIST SP 800-30~\cite{risk-standard800-30}, it introduces 
subjectivity that could affect individual technique rankings. 
Specifically, techniques for which limited empirical data is 
available may have scores that reflect the assessed severity of 
the threat more than its observed frequency. Future work should 
validate and refine these scores through structured expert surveys 
involving practitioners from across the UEFI ecosystem, including 
IBVs, OEMs, and security researchers, to reduce reliance on 
individual judgment and improve inter-rater reliability.

\noindent \textbf{Bias Toward Publicly Documented Attacks.}
The attack corpus underlying the taxonomy and risk analysis is 
necessarily limited to attacks that have been publicly disclosed. 
State-sponsored attacks, undisclosed zero-day exploits, and 
techniques employed by advanced persistent threat actors that have 
not entered the public domain are not represented. This introduces 
a systematic bias toward attacks that were either detected, 
disclosed by vendors, or deliberately made public by threat 
actors. As a consequence, the taxonomy may underrepresent 
sophisticated, covert techniques that have not yet been observed 
or reported. The risk scores of techniques associated with such 
covert activity, such as hardware supply chain compromise 
(UEFI-T0009.002), may therefore be underestimated relative to 
their true prevalence. We acknowledge this limitation and note 
that it is inherent to any empirical security taxonomy grounded 
in observed attacks rather than red-team simulation.

\noindent \textbf{Countermeasures Analysis Does Not Evaluate 
Actual Deployment.}
The countermeasures presented in Section~\ref{sec:uefi-countermeasures} 
and the applicability scores used in the effectiveness calculation 
reflect the theoretical coverage of each control against each 
attack technique. They do not account for the degree to which 
these countermeasures are actually deployed in practice, nor for 
the configuration quality of those that are deployed. For example, 
Secure Boot is widely supported but frequently misconfigured or 
disabled in real-world deployments, meaning its practical 
effectiveness is lower than its theoretical applicability score 
would suggest. A deployment-aware analysis would require empirical 
data on real-world UEFI configurations at scale, which is not 
currently available in the public domain. Future research should 
seek to close this gap through large-scale firmware analysis 
studies.

\noindent \textbf{Taxonomy Validation.}
The proposed taxonomy was derived systematically from the reviewed 
attack corpus and iteratively refined against the full set of 
documented attacks, as described in 
SubSection~\ref{sec:taxonomy_methodology}. However, it has not 
undergone formal external validation by domain experts independent 
of the authorship team. Expert validation, such as through a 
structured Delphi process or red-team evaluation, would strengthen 
confidence in the taxonomy's completeness, the appropriateness of 
technique groupings, and the accuracy of tactic boundaries. We 
present the taxonomy as a foundational framework intended to 
evolve with community input, and explicitly invite validation and 
extension by the UEFI security research community. Formal 
validation is identified as a priority direction for future work 
in Subsection~\ref{subsec:future_research_direction}.}
\section{Conclusion and Future Research}

The landscape of UEFI security has evolved dramatically over the past decade, with attacks becoming increasingly sophisticated and pervasive. Our analysis demonstrates that threat actors targeting UEFI firmware have progressed from well-resourced state actors to a broader range of malicious entities, including cybercriminal groups. This evolution is reflected by the increasing complexity of attacks, such as BlackLotus and MoonBounce, which are challenging to detect and mitigate. 

However, our risk analysis reveals a gap in the current security landscape. While several preventive measures exist---such as Secure Boot and firmware integrity checks---there is a notable lack of dynamic security mechanisms capable of adapting to the evolving threat landscape. Static defenses, while necessary, are increasingly insufficient against modern UEFI attacks. This disparity between attack sophistication and defense capabilities underscores the need for adaptive, real-time security solutions. \new{Future research should focus on creating dynamic defense mechanisms that can detect and respond to UEFI-related attack patterns in real time, as outlined below.}

\new{\subsection{Future Research Directions}\label{subsec:future_research_direction}}

\new{Our analysis reveals that the primary gaps in UEFI security are not in the absence of controls, but in their lack of adaptability, validation, and deployment assurance. Building on the taxonomy and risk analysis presented in this work, we identify several key research directions that warrant further investigation.}

\new{\noindent \textbf{Dynamic and Adaptive Defense Mechanisms.}Our risk analysis shows that high-risk techniques such as malicious payload injection, SMM exploitation, and malicious driver injection remain insufficiently mitigated despite the presence of preventive controls. This gap stems from an over-reliance on static, signature-based defenses. Future work should focus on runtime, firmware-level monitoring mechanisms capable of detecting deviations in boot flow execution, unauthorized modifications to UEFI services, and anomalous SPI flash access patterns. Such systems should operate across early boot phases, including PEI and DXE, and respond autonomously without relying on OS-level visibility.}

\new{\noindent \textbf{Automated Vulnerability Discovery in UEFI Firmware.} Reconnaissance techniques such as UEFI code analysis (UEFI-T0002) and hardware inventory (UEFI-T0003) are actively leveraged by adversaries. However, current analysis of UEFI firmware remains largely manual or limited to small-scale tools. Future research should develop scalable static and dynamic analysis pipelines tailored to firmware images, building on prior efforts such as symbolic execution and tools like efiXplorer. Particular attention is needed for third-party DXE drivers and Option ROMs, which represent a significant and under-examined attack surface.}

\new{\noindent \textbf{Supply Chain Security and Verification.} As discussed throughout the paper, the UEFI ecosystem depends on a multi-actor supply chain spanning chipset manufacturers, IBVs, ODMs, and OEMs. This distributed development and deployment model introduces multiple points of potential compromise across firmware creation, integration, update, and distribution. Future work should investigate formal verification techniques, reproducible firmware build pipelines, and cryptographic attestation mechanisms to strengthen integrity guarantees across the supply chain. Additionally, practical frameworks for integrating Software Bill of Materials (SBOM) into UEFI firmware releases remain an open challenge.}

\new{\noindent \textbf{System Management Mode (SMM) Security.} SMM exploitation (UEFI-T0033) ranks among the highest-risk techniques in our analysis, yet existing defenses remain limited, largely proprietary, and difficult to audit. Future research should explore formal isolation guarantees for SMM handlers, systematic fuzzing of SMM communication interfaces, and hardware-assisted enforcement of memory access policies to reduce the attack surface of this highly privileged execution environment.}

\new{\noindent \textbf{Emerging Technologies as Defensive Enablers.} Several nascent technologies offer promising avenues for strengthening UEFI security and warrant dedicated investigation. Hardware-based roots of trust, such as Microsoft Pluton and Google Titan, provide cryptographic attestation capabilities that are physically isolated from the SPI flash chip and may offer stronger integrity guarantees than software-only approaches. Confidential computing frameworks, including AMD SEV and Intel TDX, extend hardware-assisted isolation to firmware execution environments, potentially limiting the blast radius of SMM exploitation (UEFI-T0033) and malicious driver injection (UEFI-T0016.001). Post-quantum cryptography represents another critical frontier: as current Secure Boot implementations rely on RSA and elliptic curve schemes, future-proofing firmware signing against quantum-capable adversaries requires timely migration to quantum-resistant algorithms. Finally, machine learning-based anomaly detection applied to boot-time telemetry, such as deviations in measured boot logs or unexpected patterns in UEFI variable access, offers a path toward the adaptive, real-time defenses identified as a priority throughout this paper.}

\new{\noindent \textbf{Standardized UEFI Security Testing and Certification.} Currently, no publicly available benchmark suite exists for evaluating UEFI firmware security against known attack patterns. Developing a standardized testing and certification framework grounded in the taxonomy proposed in this paper would enable vendors to systematically assess firmware robustness. Such efforts could be analogous to existing validation programs in other domains and would provide enterprises with a consistent basis for comparing platform security.}

\new{\noindent \textbf{Regulations, Standards, and Policy Enforcement.} Because the UEFI ecosystem spans multiple stakeholders, including chipset manufacturers, IBVs, ODMs, OEMs, and enterprise operators, technical controls alone are insufficient to ensure consistent security outcomes. Future research should therefore also examine the role of regulatory requirements, certification schemes, and organizational policy frameworks in improving UEFI security. Examples include mandating secure firmware development practices, requiring supply-chain audits, enforcing coordinated vulnerability disclosure and patch management processes, and establishing minimum integrity and provenance requirements for firmware updates. Such measures can complement technical defenses by improving accountability, transparency, and baseline security assurance across the firmware lifecycle.}

\new{\noindent \textbf{Taxonomy Validation and Extension.} The MITRE ATT\&CK-like taxonomy introduced in this paper is derived from real-world attacks and proof-of-concept implementations, but has not yet undergone formal validation. Future work should involve expert-driven validation, such as Delphi studies, and red-team evaluations to refine and extend the taxonomy. In particular, emerging attack vectors targeting ARM-based platforms, virtualized firmware environments, and cloud-hosted systems should be incorporated to ensure long-term relevance.}

\new{\noindent \textbf{Operational Security and End-User Awareness.} Finally, many UEFI security mechanisms, including Secure Boot, are often misconfigured or disabled in practice. Future research should explore usable security approaches and deployment guidelines that ensure correct configuration, maintenance, and adoption of existing protections in real-world environments.}

\bibliographystyle{ACM-Reference-Format}
\bibliography{UEFISecurity}

@inproceedings{RelWorkkallenberg2014defeatingsecboot,
  title=        {Setup for failure: defeating secure boot},
  author=       {Kallenberg, Corey and Cornwell, Sam and Kovah, Xeno and Butterworth, John},
  booktitle=    {The Symposium on Security for Asia Network (SyScan)(April 2014)},
  year=         {2014}
}

@article{sarvepalli2023securing,
  title=        {Securing UEFI: An Underpinning Technology for Computing},
  author=       {Sarvepalli, Vijay},
  year=         {2023}
}

@inproceedings{RelWorkyang2020uefifuzzing,
  title=        {Uefi firmware fuzzing with simics virtual platform},
  author=       {Yang, Zhenkun and Viktorov, Yuriy and Yang, Jin and Yao, Jiewen and Zimmer, Vincent},
  booktitle=    {2020 57th ACM/IEEE Design Automation Conference (DAC)},
  pages=        {1--6},
  year=         {2020},
  organization= {IEEE}
}

@inproceedings{RelWorkrodionov2014bootkits,
  title=        {Bootkits: Past, present and future},
  author=       {Rodionov, DH Eugene and Matrosov, Aleksandr and Harley, David},
  booktitle=    {VB Conference},
  year=         {2014}
}

@inproceedings{RelWorkjiao2022uefi,
  title=        {UEFI Security Threats Introduced by S3 and Mitigation Measure},
  author=       {Jiao, Weihua and Li, Qingbao and Chen, Zhifeng and Cao, Fei},
  booktitle=    {2022 7th International Conference on Signal and Image Processing (ICSIP)},
  pages=        {734--740},
  year=         {2022},
  organization= {IEEE}
}

@article{RelWorknystrom2011uefinetworking,
  title = {UEFI Networking and Pre-OS Security},
  author = {Nyström, Magnus and Nicholes, Martin and Zimmer, Vincent J},
  journal = {Intel Technology Journal},
  volume = {15},
  number = {1},
  year = {2011}
}

@misc{uefiSPI,
  author = {{Assaf Carlsbad}},
  title = {{Moving From Common-Sense Knowledge About UEFI To Actually Dumping UEFI Firmware}},
  year = {2020},
  howpublished = {\url{https://www.sentinelone.com/labs/moving-from-common-sense-knowledge-about-uefi-to-actually-dumping-uefi-firmware/}},
  note = {Accessed: 2024-08-01}
}

@misc{eclypsiumsecureboot,
  author = {Eclypsium},
  title = {Firmware Security Realizations - Part 1: Secure Boot and DBX},
  howpublished = {\url{https://eclypsium.com/research/firmware-security-realizations-part-1-secure-boot-and-dbx/}},
  note = {Accessed: 2024-08-01}
}

@misc{mok_secboot,
  author = {Intel},
  title = {UEFI Secure Boot in Linux},
  howpublished = {\url{https://www.intel.com/content/dam/develop/external/us/en/documents/sf13-stts002-100p-820238.pdf}},
  note = {Accessed: 2024-08-01}
}

@misc{accesscontrol,
  author = {{OWASP}},
  title = {{Broken Access Control}},
  howpublished = {\url{https://owasp.org/Top10/A01_2021-Broken_Access_Control/}},
  note = {Accessed: 2024-08-01}
}

@article{zimmer2016establishing,
  title = {Establishing the Root of Trust},
  author = {Zimmer, Vincent and Krau, Michael},
  journal = {UEFI.org document dated August},
  year = {2016}
}

@inproceedings{RelWorkhagl2021secureboot,
  title = {Securing the Linux Boot Process: From Start to Finish},
  author = {Hagl, Jakob and Mann, Oliver and Pirker, Martin},
  booktitle = {Proceedings of the International Conference on Information Systems Security and Privacy (ICISSP)},
  pages = {604--610},
  year = {2021}
}

@article{wootton2016serial,
  title = {Serial Peripheral Interface (SPI)},
  author = {Wootton, Cliff and Wootton, Cliff},
  journal = {Samsung ARTIK Reference: The Definitive Developers Guide},
  pages = {335--349},
  year = {2016},
  publisher = {Springer}
}

@article{RelWorkzhou2024survey,
  title=        {A survey on the evolution of bootkits attack and defense techniques},
  author=       {Zhou, Yilin and Peng, Guojun and Li, Zichuan and Liu, Side},
  journal=      {China Communications},
  volume=       {21},
  number=       {1},
  pages=        {102--130},
  year=         {2024},
  publisher=    {IEEE}
}

@inproceedings{RelWorkkrichanov2021uefidet_mitigation,
  title=        {UEFI virtual machine firmware hardening through snapshots and attack surface reduction},
  author=       {Krichanov, Mikhail and Cheptsov, Vitaly},
  booktitle=    {2021 Ivannikov Ispras Open Conference (ISPRAS)},
  pages=        {30--36},
  year=         {2021},
  organization= {IEEE}
}

@article{RelWorkinteljounel15,
author =        {Zimmer, Vincent and Rothman, Mike},
year =          {2011},
month =         {10},
title =         {Intel Technology Journal - UEFI Today: Bootstrapping the Continuum},
volume =        {15},
journal =       {Intel Technology Journal}
}

@inproceedings{RelWorkinproceedingsTooyoung,
author =        {Bashun, Vladimir and Sergeev, Anton and Minchenkov, Victor and Yakovlev, Alexandr},
year =          {2013},
month =         {11},
pages =         {16-24},
title =         {Too young to be secure: Analysis of UEFI threats and vulnerabilities},
isbn =          {978-1-4799-4977-9},
doi =           {10.1109/FRUCT.2013.6737940}
}

@book{matrosov2019rootkits,
  title=        {Rootkits and bootkits: reversing modern malware and next generation threats},
  author=       {Matrosov, Alex and Rodionov, Eugene and Bratus, Sergey},
  year=         {2019},
  publisher=    {No Starch Press}
}

@misc{uefi_organization,
  author =       {UEFI},
  title =        {Unified Extensible Firmware Interface Forum},
  publisher =    {UEFI Forum},
  year =         2023,
  note =         {\url{https://uefi.org/specifications}}
}

@misc{uefi_specs_2_10,
  author =       {UEFI},
  title =        {Unified Extensible Firmware Interface (UEFI) Specification Release 2.10},
  publisher =    {UEFI Forum},
  year =         2022,
  edition =      {2.10},
  note =         {\url{https://uefi.org/sites/default/files/resources/UEFI_Spec_2_10_Aug29.pdf}}
}

@misc{uefi_PI_1_8,
  author =       {UEFI},
  title =        {UEFI Platform Initialization Specification},
  publisher =    {UEFI Forum},
  year =         2022,
  edition =      {1.8},
  note =         {\url{https://uefi.org/specs/PI/1.8/index.html\#}}
}

@misc{uefi_shell_spec,
  author =       {UEFI},
  title =        {UEFI Shell Specification},
  publisher =    {UEFI Forum},
  year =         2016,
  edition =      {2.2},
  note =         {\url{https://uefi.org/sites/default/files/resources/UEFI_Shell_2_2.pdf}}
}

@misc{uefi_member,
  author =       {UEFI},
  title =        {Unified Extensible Firmware Interface Forum},
  publisher =    {UEFI Forum},
  year =         2023,
  note =         {\url{https://uefi.org/about}}
}

@misc{uefi_TCG,
  author =       {TCG},
  title =        {UTCG EFI Protocol Specification},
  publisher =    {TCG},
  year =         2016,
  note =         {\url{https://trustedcomputinggroup.org/wp-content/uploads/EFI-Protocol-Specification-rev13-160330final.pdf}}
}

@misc{UEFI-MFA,
  author =      {UEFI},
  title =       {UEFI MFA},
  year =        2010,
  url =         {https://uefi.org/specs/UEFI/2.9_A/36_User_Identification.html}
}

@book{zimmer2017beyond,
  title=        {Beyond BIOS: developing with the unified extensible firmware interface},
  author=       {Zimmer, Vincent and Rothman, Michael and Marisetty, Suresh},
  year=         {2017},
  publisher=    {Walter de Gruyter GmbH \& Co KG}
}

@misc{BIOSlimitations,
  author =      {Yakup Cengiz},
  title =       {The Evolution from BIOS to UEFI},
  year =        2020,
  url =         {https://medium.com/@yakupcengiz/the-evolution-from-bios-to-uefi-cd5576f8279e},
  note =        {\url{https://medium.com/@yakupcengiz/the-evolution-from-bios-to-uefi-cd5576f8279e}}
}

@misc{mitre,
  author =      {{MITRE}},
  title =       {{MITRE ATT\&CK}},
  year =        2023,
  url =         {https://attack.mitre.org/},
}

@misc{cisa2023call,
  author =      {Jonathan Spring, Sandra Radesky},
  title =       {A Call to Action: Bolster UEFI Cybersecurity Now},
  year =        2023,
  url =         {https://www.cisa.gov/news-events/news/call-action-bolster-uefi-cybersecurity-now}
}

@misc{DarkSeaSkies,
  author =      {{Firmware Security}},
  title =       {{Wikileaks: Vault 7: Dark Matter}},
  year =        2017,
  url =         {https://firmwaresecurity.com/tag/darkmatter/}
}

@misc{SonicScrewDriver,
  author =      {{Eduard Kovacs}},
  title =       {{WikiLeaks Releases Data on CIA’s Apple Hacking Tools}},
  year =        2017,
  url =         {https://www.securityweek.com/wikileaks-releases-data-cias-apple-hacking-tools/}
}

@misc{DerStarke1,
  author =      {{Alex Matrosov}},
  title =       {{UEFI Firmware Rootkits Myths and Reality}},
  year =        2017,
  url =         {https://www.blackhat.com/docs/asia-17/materials/asia-17-Matrosov-The-UEFI-Firmware-Rootkits-Myths-And-Reality.pdf}
}

@misc{DerStarke2,
  author =      {{wikileaks}},
  title =       {{DerStarke v1.4 RC1 - IVVRR Checklist}},
  year =        2017,
  url ={https://wikileaks.org/vault7/darkmatter/document/DerStarke_v1_4_RC1_IVVRR_Checklist/}
}

@misc{DerStarke3,
  author =      {{wikileaks}},
  title =       {{DerStarke v1.4}},
  year =        2017,
  url =         {https://wikileaks.org/vault7/darkmatter/document/DerStarke_v1_4_DOC/}
}

@inproceedings{Dreamboot1,
  title={UEFI and Dreamboot},
  author={Kaczmarek, Sabastien},
  booktitle={Hack in the Box Security Conference, Kuala Lumpur, Malasia},
  year={2013}
}

@misc{Dreamboot2,
  author =      {{quarkslab}},
  title =       {{UEFI bootkit}},
  year =        2017,
  url =         {https://github.com/quarkslab/dreamboot}
}

@inproceedings{thunderstrike1,
  title={Thunderstrike: EFI firmware bootkits for Apple MacBooks},
  author={Hudson, Trammell and Rudolph, Larry},
  booktitle={Proceedings of the 8th ACM International Systems and Storage Conference},
  pages={1--10},
  year={2015}
}

@misc{thunderstrike2,
  author =      {{Hudson, Trammell }},
  title =       {{Thunderstrike}},
  year =        2014,
  url =         {https://trmm.net/Thunderstrike/}
}

@misc{darkjedi,
  author =      {{Hudson, Trammell }},
  title =       {{Thunderstrike}},
  year =        2015,
  url =         {https://trmm.net/Thunderstrike2_details/}
}

@misc{vectoredk1,
  author =      {{Takahiro Haruyama}},
  title =       {{Detecting UEFI Bootkits in the Wild (Part 1)}},
  year =        2021,
  url =         {https://blogs.vmware.com/security/2021/06/detecting-uefi-bootkits-in-the-wild-part-1.html}
}

@misc{vectoredk2,
  author =      {{hackedteam}},
  title =       {{EFI Development Kit}},
  year =        2015,
  url =         {https://github.com/hackedteam/vector-edk}
}

@inproceedings{lighteater,
  title=    {Are you giving firmware attackers a free pass},
  author=   {Kovah, Xeno and Kallenberg, Corey},
  booktitle= {Proceedings of the RSA Conference, San Francisco, CA, USA},
  pages=    {20--24},
  year=     {2015}
}

@misc{peibackdoor,
  author =      {Cr4sh},
  title =       {{PEI stage backdoor for UEFI compatible firmware}},
  year =        2022,
  publisher =   {GitHub},
  journal =     {GitHub repository},
  url =         {https://github.com/Cr4sh/PeiBackdoor}
}

@misc{thinkpwn,
  author =      {Cr4sh},
  title =       {Lenovo ThinkPad System Management Mode arbitrary code execution exploit},
  year =        2016,
  url =         {https://github.com/Cr4sh/ThinkPwn}
}

@misc{lojax1,
  author =      {ESET Research},
  title =       {LoJax: First UEFI rootkit found in the wild, courtesy of the Sednit group},
  year =        2018,
  url =         {https://www.welivesecurity.com/2018/09/27/lojax-first-uefi-rootkit-found-wild-courtesy-sednit-group/}
}

@misc{lojax2,
  author =      {hp},
  title =       {ROOTKIT TECHNICAL WHITEPAPER AN ANALYSIS OF THE MALWARE LOJAX UEFI OVERVIEW},
  year =        2021,
  url =         {https://h20195.www2.hp.com/v2/GetDocument.aspx?docname=4AA7-4019ENW}
}

@misc{lojax3,
  author =      {Trend Micro},
  title =       {LoJax UEFI Rootkit Used in Cyberespionage},
  year =        2018,
  url =         {https://www.trendmicro.com/vinfo/us/security/news/cyber-attacks/lojax-uefi-rootkit-used-in-cyberespionage}
}

@misc{mosaicregressor1,
  author =      { Eclypsium},
  title =       {PROTECT YOUR ORGANIZATION FROM MOSAICREGRESSOR AND OTHER UEFI IMPLANTS},
  year =        2020,
  url =         {https://eclypsium.com/wp-content/uploads/Protecting-Your-Organizations-From-MosaicRegressor.pdf}
}

@misc{mosaicregressor2,
  author =      {Kaspersky},
  title =       {Malware delivery through UEFI bootkit with MosaicRegressor},
  year =        2020,
  url =         {https://www.kaspersky.com/blog/mosaicregressor-uefi-malware/37252/}
}

@misc{mosaicregressor3,
  author =      {Securelist},
  title =       {MosaicRegressor: Lurking in the Shadows of UEFI},
  year =        2020,
  url =         {https://securelist.com/mosaicregressor/98849/}
}

@misc{shadowhammer1,
  author =      {Eclipsyum},
  title =       {SHADOWHAMMER AND THE FIRMWARE SUPPLY CHAIN},
  year =        2019,
  url =         {https://eclypsium.com/wp-content/uploads/ShadowHammer-and-the-Firmware-Supply-Chain.pdf}
}

@misc{shadowhammer2,
  author =      {CERT EU},
  title =       {UPDATE: Operation ShadowHammer – Compromised ASUS Computers},
  year =        2019,
  url =         {https://cert.europa.eu/publications/security-advisories/2019-007/}
}

@misc{shadowhammer3,
  author =      {Kaspersky},
  title =       {Operation ShadowHammer: new supply chain attack threatens hundreds of thousands of users worldwide},
  year =        2019,
  url =         {https://www.kaspersky.com/about/press-releases/2019_operation-shadowhammer-new-supply-chain-attack}
}

@misc{shadowhammer4,
  author =      {securelist},
  title =       {Operation ShadowHammer: a high-profile supply chain attack},
  year =        2019,
  url =         {https://securelist.com/operation-shadowhammer-a-high-profile-supply-chain-attack/90380/}
}

@misc{trickboot1,
  author =      {Eclypsium},
  title =       {TRICKBOT NOW OFFERS ‘TRICKBOOT’: PERSIST, BRICK, PROFIT},
  year =        2020,
  url =         {https://eclypsium.com/wp-content/uploads/TrickBot-Now-Offers-TrickBoot-Persist-Brick-Profit.pdf}
}

@misc{trickboot2,
  author =      {Security Affairs},
  title =       {TRICKBOOT FEATURE ALLOWS TRICKBOT BOT TO RUN UEFI ATTACKS},
  year =        2020,
  url =         {https://securityaffairs.com/111849/malware/trickbot-trickboot-uefi-attacks.html}
}

@misc{trickboot3,
  author =      {Ionut Ilascu},
  title =       {{TrickBot's new 'TrickBoot' module infects your UEFI firmware}},
  year =        2020,
  url =         {https://www.bleepingcomputer.com/news/security/trickbots-new-trickboot-module-infects-your-uefi-firmware/}
}

@misc{trickboot4,
  author =      {Ravie Lakshmanan},
  title =       {TrickBot Malware Gets UEFI/BIOS Bootkit Feature to Remain Undetected},
  year =        2020,
  url =         {https://thehackernews.com/2020/12/trickbot-malware-gets-uefibios-bootkit.html}
}

@misc{boothole1,
  author =      {Eclypsium},
  title =       {THERE'S A HOLE IN THE BOOT},
  year =        2020,
  url =         {https://eclypsium.com/blog/theres-a-hole-in-the-boot/}
}

@misc{especter1,
  author =      { Charlie Osborne},
  title =       {UEFI threats moving to the ESP: Introducing ESPecter bootkit},
  year =        2021,
  url =         {https://www.welivesecurity.com/2021/10/05/uefi-threats-moving-esp-introducing-especter-bootkit/}
}

@misc{especter2,
  author =      {Martin Smolár},
  title =       {Meet ESPecter: a new UEFI bootkit for cyber spying},
  year =        2021,
  url =         {https://www.zdnet.com/article/meet-especter-a-new-uefi-bootkit-for-cyber-spying/}
}

@misc{especter3,
  author =      {ESET},
  title =       {ESET Research discovers ESPecter, a UEFI bootkit for cyberespionage},
  year =        2021,
  url =         {https://www.eset.com/in/about/newsroom/press-releases/research/eset-research-discovers-especter-a-uefi-bootkit-for-cyberespionage/}
}

@misc{finspy1,
  author =      {SecureList},
  title =       {FinSpy: unseen findings},
  year =        2021,
  url =         {https://securelist.com/finspy-unseen-findings/104322/}
}

@misc{finspy2,
  author =      {Kaspersky},
  title =       {FinFisher spyware improves its arsenal with four levels of obfuscation, UEFI infection and more},
  year =        2021,
  url =         {https://www.kaspersky.com/about/press-releases/2021_finfisher-spyware-improves-its-arsenal-with-four-levels-of-obfuscation-uefi-infection-and-more}
}

@misc{finspy3,
  author =      {Eclypsium},
  title =       {FINSPY UEFI AND MBR BOOTKIT},
  year =        2021,
  url =         {https://eclypsium.com/blog/finspy-uefi-and-mbr-bootkit/}
}

@misc{moonbounce1,
  author =      {Bill Toulas},
  title =       {New MoonBounce UEFI malware used by APT41 in targeted attacks},
  year =        2022,
  url =         {https://www.bleepingcomputer.com/news/security/new-moonbounce-uefi-malware-used-by-apt41-in-targeted-attacks/}
}

@misc{moonbounce2,
  author =      {MARK LECHTIK},
  title =       {MoonBounce: the dark side of UEFI firmware},
  year =        2022,
  url =         {https://securelist.com/moonbounce-the-dark-side-of-uefi-firmware/105468/}
}

@misc{cosmicstrand1,
  author =      {Eclypsium},
  title =       {YET ANOTHER UEFI BOOTKIT DISCOVERED: MEET COSMICSTRAND},
  year =        2022,
  url =         {https://eclypsium.com/blog/yet-another-uefi-bootkit-discovered-meet-cosmicstrand/}
}

@misc{cosmicstrand2,
  author =      {Secure List},
  title =       {CosmicStrand: the discovery of a sophisticated UEFI firmware rootkit},
  year =        2022,
  url =         {https://securelist.com/cosmicstrand-uefi-firmware-rootkit/106973/}
}

@misc{cosmicstrand3,
  author =      {Julia Glazova},
  title =       {CosmicStrand: a UEFI rootkit},
  year =        2022,
  url =         {https://www.kaspersky.com/blog/cosmicstrand-uefi-rootkit/45017/}
}

@misc{cosmicstrand4,
  author =      {Tara Seals},
  title =       {Rare CosmicStrand UEFI Rootkit Swings into Cybercrime Orbit},
  year =        2022,
  url =         {https://www.darkreading.com/endpoint-security/rare-cosmicstrand-uefi-rootkit-cybercrime-orbit}
}

@misc{blacklotus1,
  author =      {Martin Smolár},
  title =       {BlackLotus UEFI bootkit: Myth confirmed},
  year =        2022,
  url =         {https://www.welivesecurity.com/2023/03/01/blacklotus-uefi-bootkit-myth-confirmed/}
}

@book{thompson2003BIOS,
  title=        {PC hardware in a nutshell: a desktop quick reference},
  author=       {Thompson, Robert Bruce and Thompson, Barbara Fritchman},
  year=         {2003},
  publisher=    {" O'Reilly Media, Inc."}
}

@inproceedings{wilkins2013uefi,
  title=        {UEFI secure boot in modern computer security solutions},
  author=       {Wilkins, Richard and Richardson, Brian},
  booktitle=    {UEFI forum},
  pages=        {1--10},
  year=         {2013}
}

@misc{UEFIvarPkKekDbDbx,
  author =      {{Intel}},
  title =       {{UEFI Security and Networking Advancements}},
  year =        2011,
  url =         {https://www.intel.com/content/dam/develop/external/us/en/documents/sf11-efis001-100-820238.pdf}
}

@misc{eclypsium2024kev,
  author       = {Eclypsium},
  title        = {What You Need to Know About the Latest KEV Updates},
  howpublished = {\url{https://eclypsium.com/blog/what-you-need-to-know-about-the-latest-kev-updates/}},
  year         = {2024},
  note         = {Accessed: 2024-08-08}
}

@misc{uefi_firmware_security_concerns, 
author = {UEFI Forum}, 
title = {UEFI Firmware - Security Concerns and Best Practices}, 
url = {https://uefi.org/sites/default/files/resources/UEFI%20Firmware%20-%20Security%20Concerns%20and%20Best%20Practices.pdf}, 
year = {n.d.} 
}

@misc{nsa_uefi_defensive_practices, 
author = {National Security Agency (NSA)}, 
title = {UEFI Defensive Practices Guidance}, 
note = {\url{https://www.nsa.gov/portals/75/documents/what-we-do/cybersecurity/professional-resources/ctr-uefi-defensive-practices-guidance.pdf}}, 
year = {n.d.} 
}

@misc{defense_secure_boot_customization, 
author = {Department of Defense}, 
title = {CTR-UEFI Secure Boot Customization}, 
note = {\url{https://media.defense.gov/2023/Mar/20/2003182401/-1/-1/0/CTR-UEFI-SECURE-BOOT-CUSTOMIZATION-20230317.PDFnote}}, 
year = {2023} 
}

@misc{uefi_secure_boot_insyde, 
author = {UEFI Forum}, 
title = {UPFS11 P2 SecureBoot Insyde}, 
note = {\url{https://uefi.org/sites/default/files/resources/UPFS11_P2_SecureBoot_Insyde.pdf}},
year = {n.d.} 
}

@misc{eclypsium_spi_write_protections, 
author = {Eclypsium}, 
title = {Firmware Security Realizations Part 3: SPI Write Protections}, 
note = {\url{https://eclypsium.com/research/firmware-security-realizations-part-3-spi-write-protections/}}, 
year = {n.d.} 

}

@misc{uefi_virtual_summit_2022, 
author = {UEFI Forum}, 
title = {UEFI Virtual Summit 2022: Tackling Security Through the Supply Chain}, 
note = {\url{https://uefi.org/sites/default/files/resources/UEFI_VirtualSummit_2022_Tackling_Security_Through_the_Supply_Chain_Draft_Final_v4.pdf}}, 
year = {2022} 
}

@misc{blackhat_uefi_safeguarding, 
author = {Blackhat}, 
title = {Safeguarding UEFI Ecosystem Firmware Supply Chain is Hardcoded}, 
note = {\url{https://i.blackhat.com/USA21/Wednesday-Handouts/us-21-Safeguarding-UEFI-Ecosystem-Firmware-Supply-Chain-Is-Hardcoded.pdf}}, 
year = {2021} 
}

@misc{SBOM-supplychain,
  title =       {The Various Shades of Supply Chain: SBOM, N-Days and Zero Trust},
  author =      {Alex Matrosov, Richard Hughes, Kai Michaelis },
  year =        {2023},
  howpublished = {Available online at: \url{https://www.youtube.com/watch?v=DwCgPjvssD4}},
  note =        {\url{https://www.blackhat.com/asia-23/briefings/schedule/\#the-various-shades-of-supply-chain-sbom-n-days-and-zero-trust-31253}},
}

@misc{qual-risk,
  author = {ISACA},
  title = {Risk Assessment and Analysis Methods: Qualitative and Quantitative},
  year = {2021},
  howpublished = {\url{https://www.isaca.org/resources/isaca-journal/issues/2021/volume-2/risk-assessment-and-analysis-methods}},
  note = {Accessed: 2024-08-01}
}

@misc{uefi_sbom_support, 
author = {UEFI Forum}, 
title = {UEFI Support for Software Bill of Materials (SBOM)}, 
url = {ttps://uefi.org/sites/default/files/resources/UEFI%20Support%20for%20Software%20Bill%20of%20Materials%20%28SBOM%29_AMI_UEFI_Final_Deck.pdf}, 
year = {n.d.} 
}

@misc{uefi_plugfest_2012, 
author = {UEFI Forum}, 
title = {UEFI Plugfest 2012Q1 v3 AMI}, 
note = {\url{https://uefi.org/sites/default/files/resources/UEFI_Plugfest_2012Q1_v3_AMI.pdf}}, 
year = {2012}
}

@misc{spiceworks_version_control, 
author = {Spiceworks}, 
title = {What is Version Control?}, 
note = {\url{https://www.spiceworks.com/tech/devops/articles/what-is-version-control/}}, 
year = {n.d.} 
}

@misc{rapid7_vulnerability_management, 
author = {Rapid7}, 
title = {Vulnerability Management Program Framework}, 
note = {\url{https://www.rapid7.com/fundamentals/vulnerability-management-program-framework/}}, 
year = {n.d.} 
}

@misc{defense_signed_software_execution, 
author = {Department of Defense}, 
title = {Enforce Signed Software Execution Policies}, 
url = {https://media.defense.gov/2019/Sep/09/2002180334/-1/-1/0/Enforce%20Signed%20Software%20Execution%20Policies%20-%20Copy.pdf}, 
year = {2019} 
}

@misc{uefi_memory_protection, 
author = {UEFI Forum}, 
title = {Hardening the Core Enhanced Memory Protection}, 
url = {https://uefi.org/sites/default/files/resources/Hardening%20the%20Core%20Enhanced%20Memory%20Protection_Beebe.pdf}, 
year = {n.d.}
}

@misc{entitle_mfa, 
author = {Entitle}, 
title = {Multi-Factor Authentication (MFA)},
note = {\url{https://www.entitle.io/resources/glossary/multi-factor-authentication-mfa}}, 
year = {n.d.} 
}

@misc{uefi_kms_implementation, 
author = {UEFI Forum}, 
title = {Implementing and Using the UEFI KMS}, 
url = {https://uefi.org/sites/default/files/resources/Implementing%20and%20Using%20the%20UEFI%20KMS_AMI_Final.pdf}, 
year = {n.d.} 
}

@misc{uefi_key_management_service, 
author = {UEFI Forum}, 
title = {UEFI Key Management Service (KMS) With TPM}, 
url = {https://uefi.org/sites/default/files/resources/UEFI%20Key%20Management%20Service%20%28KMS%29%20With%20TPM%C2%A0_Otumfuor%20and%20Polyudov.pdf}, 
year = {n.d.} 
}

@misc{uefi_plugfest_2014, 
author = {UEFI Forum}, 
title = {UEFI Plugfest 2014 Phoenix}, 
note = {\url{https://uefi.org/sites/default/files/resources/2014_UEFI_Plugfest_06_Phoenix.pdf}},
year = {2014} 
}

@misc{uefi_intel_defense, 
author = {UEFI Forum}, 
title = {Intel Attacking and Defending the Platform}, 
url = {https://uefi.org/sites/default/files/resources/Intel_Attacking%20and%20Defending%20the%20Platform.pdf}, 
year = {n.d.} 
}

@misc{uefi_secure_boot_driver_signing, 
author = {UEFI Forum}, 
title = {Secure Boot and Driver Signing}, 
note = {\url{https://uefi.org/specs/UEFI/2.9_A/32_Secure_Boot_and_Driver_Signing.html}}, 
year = {n.d.} 
}

@misc{uefi_secure_coding_firmware, 
author = {UEFI Forum}, 
title = {UEFI March Webinar Secure Coding for UEFI Firmware}, 
url = {https://uefi.org/sites/default/files/resources/UEFI%20March%20Webinar%20Secure%20Coding%20for%20UEFI%20Firmware.pdf}, 
year = {n.d.} 
}

@misc{eclypsium_evil_maid_attacks, 
author = {Eclypsium}, 
title = {Evil Maid Firmware Attacks Using USB Debug},
note = {\url{https://eclypsium.com/research/evil-maid-firmware-attacks-using-usb-debug/}}, 
year = {n.d.} 
}

@misc{cisa_cybersecurity_response_playbooks, 
author = {Cybersecurity and Infrastructure Security Agency (CISA)}, 
title = {Federal Government Cybersecurity Incident and Vulnerability Response Playbooks},
note = {\url{https://www.cisa.gov/sites/default/files/2024-03/Federal_Government_Cybersecurity_Incident_and_Vulnerability_Response_Playbooks_508C.pdf}}, 
year = {2024} 
}

@misc{nist_sp800_147, 
author = {National Institute of Standards and Technology (NIST)}, 
title = {NIST Special Publication 800-147: BIOS Protection Guidelines},
note = {\url{https://nvlpubs.nist.gov/nistpubs/legacy/sp/nistspecialpublication800-147.pdf}},
year = {2011} 
}

@misc{paloalto_glupteba_malware,
author = {Palo Alto},
title = {Diving Into Glupteba's UEFI Bootkit},
note = {\url{https://unit42.paloaltonetworks.com/glupteba-malware-uefi-bootkit/}},
year = {n.d.}
}

@misc{cisa_uefi_cybersecurity_call_to_action, 
author = {Cybersecurity and Infrastructure Security Agency (CISA)}, 
title = {Call to Action: Bolster UEFI Cybersecurity Now}, 
note = {\url{https://www.cisa.gov/news-events/news/call-action-bolster-uefi-cybersecurity-now}}, 
year = {n.d.} 
}

@misc{labunets_efixplorer, 
author = {Labunets, Andrei}, 
title = {efiXplorer: Hunting For UEFI Firmware Vulnerabilities At Scale With Automated Static Analysis}, 
note = {\url{https://i.blackhat.com/eu-20/Wednesday/eu-20-Labunets-efiXplorer-Hunting-For-UEFI-Firmware-Vulnerabilities-At-Scale-With-Automated-Static-Analysis.pdf}},
year = {2020} 
}

@misc{binarly_symbolic_execution,
author = {Binarly}, 
title = {Using Symbolic Execution to Detect UEFI Firmware Vulnerabilities},
note = {\url{https://www.binarly.io/blog/using-symbolic-execution-to-detect-uefi-firmware-vulnerabilities}},
year = {n.d.}
}

@misc{eclypsium_dma_attacks,
author = {Eclypsium}, 
title = {Direct Memory Access Attacks: A Walk Down Memory Lane}, 
note = {\url{https://eclypsium.com/research/direct-memory-access-attacks-a-walk-down-memory-lane/}}, 
year = {n.d.} 
}

@misc{eclypsium_sunburst_firmware_attack, 
author = {Eclypsium}, 
title = {Sunburst Firmware Supply Chain Attack IR},
note = {\url{https://eclypsium.com/blog/sunburst-firmware-supply-chain-attack-ir/}}, 
year = {n.d.} 
}

@misc{microsoft_boot_manager_revocations, 
author = {Microsoft},
title = {KB5025885: How to Manage the Windows Boot Manager Revocations for Secure Boot Changes Associated with CVE-2023-24932}, 
note = {\url{https://support.microsoft.com/en-us/topic/kb5025885-how-to-manage-the-windows-boot-manager-revocations-for-secure-boot-changes-associated-with-cve-2023-24932-41a975df-beb2-40c1-99a3-b3ff139f832d}}, 
year = {2023}
}

@misc{uefi_secure_boot_modern_security, 
author = {UEFI Forum},
title = {UEFI Secure Boot in Modern Computer Security Solutions}, 
note = {\url{https://uefi.org/sites/default/files/resources/UEFI_Secure_Boot_in_Modern_Computer_Security_Solutions_2013.pdf}}, 
year = {2013}
}

@misc{uefi_vulnerability_management, 
author = {UEFI Forum}, 
title = {Vulnerability Management in UEFI}, 
url = {https://uefi.org/sites/default/files/resources/Vulnerability%20Management%20in%20UEFI_Mullen.pdf},
year = {n.d.}
}

@article{IOMMU_DMA,
  title=        {A tour beyond BIOS: Using IOMMU for DMA protection in UEFI firmware},
  author=       {Yao, Jiewen and Zimmer, Vincent J and Zeng, Star},
  journal=      {Intel Corporation},
  volume=       {1497},
  year=         {2017}
}

@misc{bitlocker,
  author =      {Vinay Pamnani},
  title =       {{BitLocker Countermeasures}},
  year =        2023,
  publisher =   {GitHub},
  journal =     {Microsoft Learning Center},
  url =         {https://learn.microsoft.com/en-us/windows/security/operating-system-security/data-protection/bitlocker/bitlocker-countermeasures\#uefi-and-secure-}
}

@article{update-runtime-exploitation,
  title=        {Extreme privilege escalation on Windows 8/UEFI systems},
  author=       {Kallenberg, Corey and Kovah, Xeno and Butterworth, John and Cornwell, Sam},
  journal=      {BlackHat, Las Vegas, USA},
  year=     {2014}
}

@inproceedings{SMMrootkit,
  title=        {The design of the simple SMM rootkit},
  author=       {Szczypiorski, Krzysztof and Szaknis, Micha{\l}},
  booktitle=    {9th International Conference on Wireless Communication and Sensor Networks},
  year=         {2022},
  organization= {Association for Computing Machinery}
}

@misc{flashdescriptor,
  author = {Xeno Kovah and Corey Kallenberg},
  title = {Advanced x86: BIOS and System Management Mode Internals Flash Descriptor},
  howpublished = {\url{https://opensecuritytraining.info/IntroBIOS_files/Day2_02_Advanced\%20x86\%20-\%20BIOS\%20and\%20SMM\%20Internals\%20-\%20Flash\%20Descriptor.pdf}},
  note = {Accessed: 2024-08-01}
}

@misc{best-practice-sig-sec-devp,
  author = {Cloud Security Industry Summit - Supply Chain Technical Working Group},
  title = {Secure Firmware Development Best Practices},
  howpublished = {\url{https://www.opencompute.org/documents/csis-firmware-security-best-practices-position-paper-version-1-0-pdf}},
  note = {Accessed: 2024-08-01}
}

@misc{risk-standard800-30,
  author =      {NIST},
  title =       { NIST Special Publication 800-30     Revision 1 },
  url =         {https://nvlpubs.nist.gov/nistpubs/Legacy/SP/nistspecialpublication800-30r1.pdf},
  note =        {\url{https://nvlpubs.nist.gov/nistpubs/Legacy/SP/nistspecialpublication800-30r1.pdf}}
}

@inproceedings{CAR,
  title=        {Attacking Intel UEFI by Using Cache Poisoning},
  author=       {Wang, Dong and Dong, Wei Yu},
  booktitle=    {Journal of Physics: Conference Series},
  volume=       {1187},
  number=       {4},
  pages=        {042072},
  year=         {2019},
  organization= {IOP Publishing}
}

@techreport{nist,
  title     = {Cybersecurity White Paper: Securing Property Using Property-Based Attestation},
  author    = {{National Institute of Standards and Technology}},
  year      = {2020},
  institution = {National Institute of Standards and Technology},
  address   = {Gaithersburg, MD},
  url       = {https://nvlpubs.nist.gov/nistpubs/CSWP/NIST.CSWP.29.pdf},
  note      = {Accessed: August 21, 2024},
}

\newpage

\end{document}